\documentclass[prd,twocolumn,showpacs,preprintnumbers,nofootinbib,superscriptaddress,amsmath,amssymb,
aps]{revtex4} 
\usepackage{mathrsfs}
\usepackage{graphicx}% Include figure files
\usepackage{dcolumn}% Align table columns on decimal point
\usepackage{bm}% bold math
\renewcommand{\labelenumi}{(\roman{enumi})}
\usepackage{comment}
\usepackage{color}
\usepackage{soul}

\begin{document}

\preprint{CHIBA-EP-239, 2019.05.13}

\title{Complex poles and spectral function of Yang-Mills theory
}% Force line breaks with \\
% \thanks{A footnote to the article title}

\author{Yui Hayashi}
\email{yhayashi@chiba-u.jp}
\affiliation{
Department of Physics, Graduate School of Science and Engineering, Chiba University, Chiba 263-8522, Japan
}

\author{Kei-Ichi Kondo}
\email{kondok@faculty.chiba-u.jp}
\affiliation{
Department of Physics, Graduate School of Science and Engineering, Chiba University, Chiba 263-8522, Japan
}
\affiliation{
Department of Physics, Graduate School of Science, Chiba University, Chiba 263-8522, Japan
}


\begin{abstract}
We derive general relationships between the number of complex poles of a propagator and the sign of the spectral function originating from the branch cut in the Minkowski region under some assumptions on the asymptotic behaviors of the propagator. 
We apply this relation to the mass-deformed Yang-Mills model with one-loop quantum corrections, which is identified with a low-energy effective theory of the Yang-Mills theory, to show that the gluon propagator in this model has a pair of complex conjugate poles or ``tachyonic'' poles of multiplicity two, in accordance with the fact that the gluon field has a negative spectral function, while the ghost propagator has at most one ``unphysical'' pole.
%Moreover, we show that some other cases available in the literature are consistent with the general relation.
Finally, we discuss implications of these results for gluon confinement and other nonperturbative aspects of the Yang-Mills theory.

\end{abstract}

\maketitle

%\tableofcontents

\section{INTRODUCTION}

Color confinement is one of the central features of the strong interactions, which implies that colored particles are absent in the observed spectrum. 
In the last decade, numerous studies have focused on evaluating the infrared (IR) behaviors of gluon and ghost propagators in the Landau gauge of the Yang-Mills theory. Consequently, the so-called \textit{decoupling solution} got to be supported instead of the so-called \textit{scaling solution} as the confining solution of the Yang-Mills theory in the space-time dimension $D=4$ based on analytical studies \cite{decoupling-analytical} and numerical simulations on the lattice \cite{decoupling-lattice, DOS16}. 
An important feature of the decoupling solution lies in a fact that the running gauge coupling remains finite for nonvanishing momentum and eventually goes to zero in the limit of vanishing momentum, which cannot be derived from the standard perturbation theory which predicts that the running gauge coupling diverges at a certain momentum, the infrared Landau pole.

This issue in the Landau gauge Yang-Mills theory was addressed by various continuum approaches.
Among them, the massive extension of the Yang-Mills theory in the Landau gauge \cite{TW10,TW11}, which we call the \textit{massive Yang-Mills model} for short, has succeeded to reproduce the results of numerical simulations on the lattice with good accuracy in the strict one-loop level, and also provide both decoupling and scaling solutions without the infrared Landau pole by taking into account the renormalization group improvement \cite{TW10,TW11,RSTW17,ST12}. 
Indeed, the perturbative analysis and its systematic improvement in the massive Yang-Mills model is relatively easier than solving the equations derived from more elaborate nonperturbative approaches, e.g., the functional renormalization group \cite{CFMPS16,CPRW18} and the Schwinger-Dyson equation \cite{SFK12}.
The massive Yang-Mills model is regarded as a low-energy effective theory of the original Yang-Mills theory and also has been successfully applied to the finite temperature Yang-Mills theory \cite{RSTW14-17}, see also \cite{Kondo15}. 
Similar ideas of massive gluons have been employed in \cite{quasi-particle_massive_gluon}, phenomenological models of the deconfined phase.

In the original works \cite{TW10,TW11} the massive Yang-Mills model in the Landau gauge was identified with a special parameter limit of the Curci-Ferrari model \cite{CF76b}. 
However, the Curci-Ferrari model is not invariant under the usual Becchi-Rouet-Stora-Tyutin (BRST) transformation, but invariant just under the modified BRST transformation which does not respect the usual nilpotency. 
Recently, another identification of the massive Yang-Mills model has been given as a reduction of the gauge-invariantly extended model where the gauge-invariant mass term is provided through the gauge-independent Brout-Englert-Higgs mechanism \cite{Kondo16,Kondo18} from the complementary gauge-scalar model with a radially fixed scalar field, see \cite{KSOMH18} for more details.

A large number of previous works were so far devoted to investigating the propagators in the Euclidean space to obtain the results which are compatible with the data of numerical simulations on the lattice.
On the other hand, several phenomenological models \cite{Gribov78,Stingl86,Zwanziger90,DGSVV2008,BDGHSVZ10} predict the existence of complex poles in the gluon propagator.
If this is true, we must seriously consider the complex structure of the propagator.
Indeed, the existence of such complex poles invalidates the usual form of the spectral representation of the K\"all\'en-Lehmann type~\cite{spectral_repr_UKKL}. 
The detailed considerations on the complex structure will provide useful information to figure out the nonperturbative aspects of the theory and play pivotal roles to obtain a clear description of confinement in the Yang-Mills theory.
In the present paper, we pay special attention to complex poles and the branch cut of the gluon and ghost propagators in the Landau gauge Yang-Mills theory. 
First, we derive general relations which connect the number of complex poles and zeros of a propagator to the sign (positivity or negativity) of the spectral function originating from discontinuity across the branch cut in the Minkowski region
by assuming the asymptotic behaviors of the propagator in the neighborhood of the origin and the region far from the origin in the complex momentum plane.
Then we apply one of these relations to the massive Yang-Mills model with one-loop quantum corrections to show that the gluon propagator has a pair of complex conjugate poles or ``tachyonic'' poles (in the Euclidean region) of multiplicity two, as a consequence of the fact that the gluon field has a negative spectral function in the one-loop level. 
On the other hand, we show that the ghost propagator has no complex pole and has at most one ``unphysical'' pole (besides the Minkowski region). 
Moreover, we investigate the pole structure of the gluon propagator in the whole parameter space of the massive Yang-Mills model to examine the possible crossover from confining region to Higgs-like region in the single confined phase, 
which has been recently discussed in \cite{CFMPS16,RSTW17} and \cite{KSOMH18} from the different viewpoints.

This paper is organized as follows. 
In Sec.~II, we first review the well-known spectral representation of a propagator from the viewpoint of analyticity (holomorphy) and subsequently consider the generalization in the presence of complex poles.
In the presence of a pair of complex conjugate poles, in particular, we derive a modified form of the superconvergence relation for a spectral function. 
In Sec.~III, we show by using the argument principle how the number of complex poles of a propagator is determined under assumptions on the sign of the spectral function and the asymptotic behaviors of the propagator at the small and large momenta in the complex momentum plane. 
In Sec.~IV, we introduce the massive Yang-Mills model and summarize the result of the one-loop quantum corrections in this model. 
In Sec.~V, we reveal the pole structure of the gluon and ghost propagators in the massive Yang-Mills model within the one-loop level, in the light of the general consideration given in Sec.~III.
First, we show that the spectral function is always negative in the massive Yang-Mills model to one-loop order. 
Then we establish that the gluon propagator has a pair of complex conjugate poles and that the ghost propagator has no complex poles in this model. 
The final Sec.~VI is devoted to conclusions and discussions.
We discuss implications of the pole structure for the crossover between confining and Higgs-like regions in the parameter space and possible effects of the renormalization group improvement qualitatively. 
In Appendix A, we discuss various cases of the relation between poles and the winding number.
In Appendix B, we analyze the other models available in the literature to show the consistency with the general relation.

\section{GENERALIZATION OF THE SPECTRAL REPRESENTATION}
In this section, we rederive the spectral representation of the K\"all\'en-Lehmann form for a propagator and its generalization in light of analyticity. As an important example of this generalization, in particular, we consider a propagator with one pair of complex conjugate poles.
\subsection{Analyticity}
First, we consider a propagator $D(k^2)$ as a complex function of the complex variable $z = k^2$ with the following properties \cite{Kondo03}: 
\begin{enumerate}
 \item $D(z)$ is holomorphic except singularities on the positive real axis.
 \item $D(z) \rightarrow 0$ as $|z| \rightarrow \infty$.
 \item $D(z)$ is real on the negative real axis.
\end{enumerate}
In this case, the propagator $D(k^2)$ has the spectral representation of the K\"all\'en-Lehmann form. It is well known that a propagator takes the K\"all\'en-Lehmann form under the assumptions of the spectral condition, Poincar\'e invariance and the completeness of the state space \cite{spectral_repr_UKKL}.

\begin{figure}[tbp]
 \begin{center}
 \includegraphics[width=70mm]{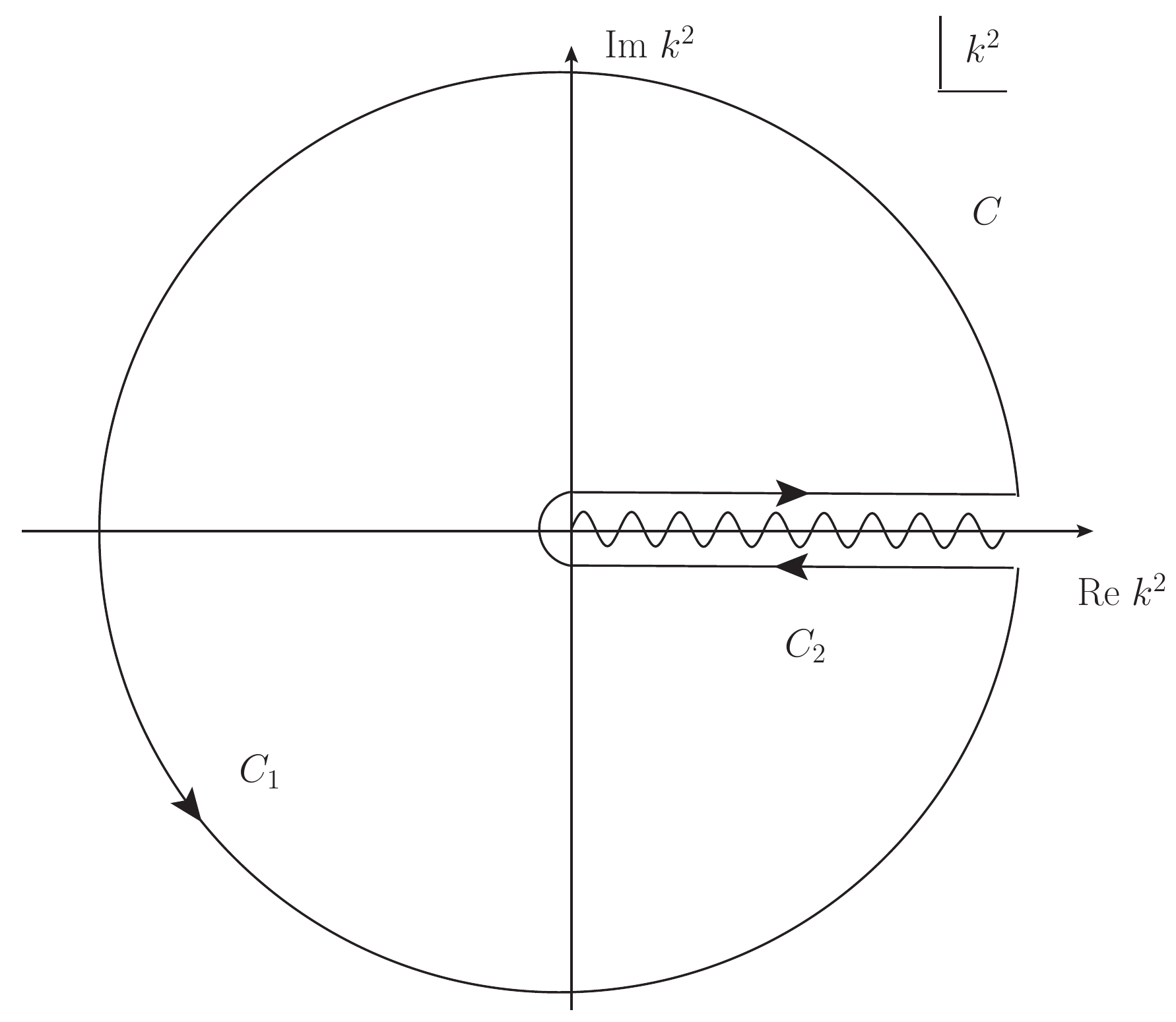}
 \end{center}
 \caption{Contour $C$ on the complex $k^2$ plane circumventing singularities of a propagator $D(k^2)$ on the positive real axis. The propagator $D(k^2)$ is holomorphic inside $C$. The contour $C$ consists of the large circle $C_1$ and the path $C_2$ winding around the positive real axis.}
 \label{fig:section2_analiticity.pdf}
\end{figure}

Then we choose a closed loop $C$ (shown in Fig.~\ref{fig:section2_analiticity.pdf}) so that $D(z)$ is holomorphic in the interior of the region with the boundary $C$.
The Cauchy integral formula and the first assumption (i) yield
\begin{align}
 D(k^2) = \frac{1}{2 \pi i} \oint_C d \zeta \frac{D(\zeta)}{\zeta - k^2}. \label{eq:Cauchy_formula_A}
\end{align}
The contour $C$ is divided into two paths $C=C_1+C_2$ as depicted in Fig.~1, where $C_1$ is the sufficiently large circle and $C_2$ is the path wrapping around the branch cut on the real positive axis.
From the second assumption (ii), the integration along $C_1$ vanishes. Then the expression (\ref{eq:Cauchy_formula_A}) can be cast into
\begin{align}
 D(k^2) &= \frac{1}{2 \pi i} \int_{C_2} d \zeta \frac{D(\zeta)}{\zeta - k^2} \notag \\
 &= \frac{1}{2 \pi i} \int_0 ^\infty d \zeta \frac{D(\zeta+i\epsilon) - D(\zeta - i \epsilon)}{\zeta - k^2}.
\end{align}
Moreover, because of the third assumption (iii), the Schwarz reflection principle follows
\begin{align}
D(z^*) = [D(z)]^*.
\end{align}
Therefore, we have
\begin{align}
 D(k^2) &= \frac{1}{\pi} \int_0 ^\infty d \zeta \frac{{\rm Im}\ D(\zeta+i\epsilon)}{\zeta - k^2}.
\end{align}
 Finally, $D(k^2)$ has the spectral representation of the K\"all\'en-Lehmann type with the spectral function $\rho$,
\begin{align}
 D(k^2)&= \int_0 ^\infty d \sigma^2 \frac{\rho(\sigma^2)}{\sigma^2 - k^2}, \label{eq:KL_spectral_repr}\\
 \rho(\sigma^2) &:= \frac{1}{\pi} {\rm Im}\ D(\sigma^2+i\epsilon).
\end{align} 

\subsection{Complex poles}
Second, we consider a propagator $D(k^2)$ with the following properties \cite{Siringo17a}:
\begin{enumerate}
 \item $D(z)$ is holomorphic except singularities on the positive real axis and a finite number of simple poles.
 \item $D(z) \rightarrow 0$ as $|z| \rightarrow \infty$.
 \item $D(z)$ is real on the negative real axis.
\end{enumerate}
This is a straightforward generalization of the previous case, allowing the existence of complex simple poles. 
%[[***Such singularities can appear in unphysical particles whose bound states are physical \cite{BDGHSVZ10}.]]

\begin{figure}[tbp]
 \begin{center}
 \includegraphics[width=70mm]{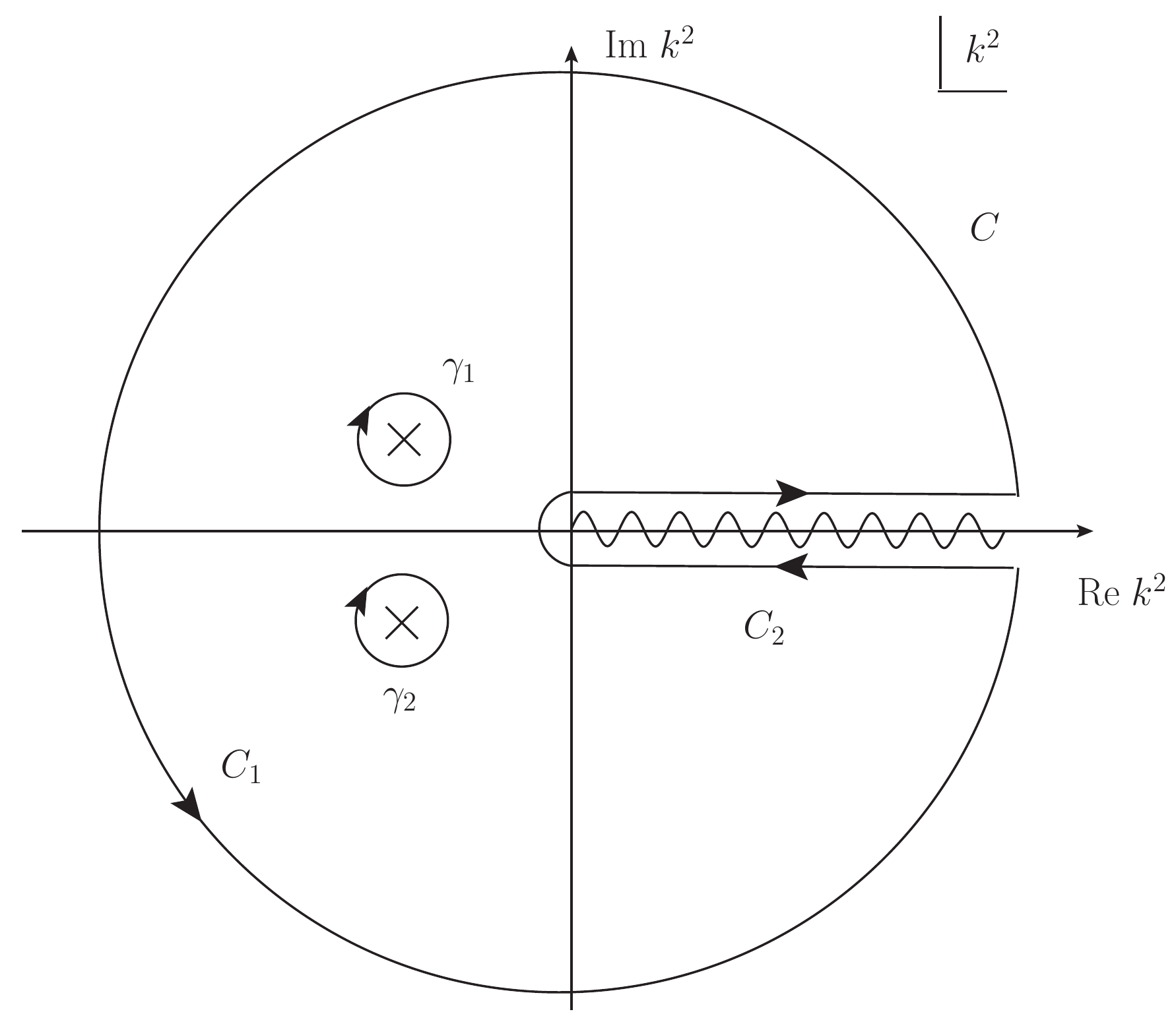}
 \end{center}
 \caption{Contour $\tilde{C}$ on the complex $k^2$ plane avoiding complex poles and the singularities on the positive real axis. The contour $\gamma_\ell$ surrounds the pole $z_\ell$ clockwise. The propagator $D(k^2)$ is holomorphic in the region bounded by the contour $\tilde{C} = C \cup \{ \gamma_\ell \}_{\ell = 1} ^n$.}
 \label{fig:section2_complex.pdf}
\end{figure}

Suppose that the propagator has simple complex poles at $z = z_\ell (\ell = 1, \cdots, n)$. We choose the contour $\tilde{C}$ so that $D(z)$ is holomorphic inside $\tilde{C} = C \cup \{ \gamma_\ell \}_{\ell = 1} ^n$ as shown in Fig.~\ref{fig:section2_complex.pdf}, where $\gamma_\ell$ is a small contour oriented clockwise around $z_\ell$.
Then the Cauchy integral formula gives
\begin{align}
 D(k^2) = \frac{1}{2 \pi i} \oint_C d \zeta \frac{D(\zeta)}{\zeta - k^2} + \sum_{\ell=1}^n \frac{1}{2 \pi i} \oint_{\gamma_\ell} d \zeta \frac{D(\zeta)}{\zeta - k^2} .
\end{align}
Notice that the integral along $C_1$ vanishes as the previous one. Then we have the generalized spectral representation,
\begin{align}
 D(k^2) &= \int_0 ^\infty d \sigma^2 \frac{\rho(\sigma^2)}{\sigma^2 - k^2} + \sum_{\ell=1}^n \frac{Z_\ell}{z_\ell - k^2}, \label{eq:spec_repr_complex} \\
 \rho(\sigma^2) &:= \frac{1}{\pi} {\rm Im}\ D(\sigma^2+i\epsilon), \label{eq:dispersion_complex} \\ 
 Z_\ell &:= \oint_{\gamma_\ell} \frac{d k^2}{2 \pi i} D(k^2). \label{eq:dispersion_residue}
\end{align}
If the poles were not simple, the second term would be modified as
\begin{align}
& \sum_{\ell=1}^n \frac{1}{2 \pi i} \oint_{\gamma_\ell} d \zeta \frac{D(\zeta)}{\zeta - k^2} = \sum_{\ell=1}^n \sum_{n=1}^\infty \frac{a_{-n}^\ell}{(k^2 - z_\ell)^n}, \notag \\
&a_{-n}^\ell = - \frac{1}{2 \pi i} \oint_{\gamma_\ell} d k^2 D(k^2) (k^2 - z_\ell)^{n-1}.
\end{align}
Note that the poles must appear as real poles or pairs of complex conjugate poles as a consequence of the Schwarz reflection principle $D(z^*) = [D(z)]^*$.

\subsection{One pair of complex poles}
Finally, we focus on a propagator with a pair of complex conjugate simple poles, which is an important case. For example,
the propagator consisting of a pair of complex conjugate poles reproduces well the results of the numerical simulations \cite{Gribov78,Stingl86,Zwanziger90,DGSVV2008,BDGHSVZ10}. In addition, the gluon propagator of the massive Yang-Mills model exhibits one pair of complex conjugate poles as will be shown in Sec.~V.

Let us see novel aspects of a propagator with one pair of complex conjugate simple poles at $k^2 = v \pm i w$, where we can set $w > 0$ without loss of generality. In this case, the generalized spectral representation (\ref{eq:spec_repr_complex}) reduces to,
\begin{align}
 D(k^2) &= \int_0 ^\infty d \sigma^2 \frac{\rho(\sigma^2)}{\sigma^2 - k^2} \notag \\
 & ~~~ + \frac{Z}{(v+iw) - k^2} + \frac{Z^*}{(v-iw) - k^2}. \label{eq:one_pair_complex}
\end{align}
We now assume that the propagator has the following asymptotic behavior in the region far from the origin of the complex $k^2$ plane,
\begin{align}
\lim_{|k^2| \rightarrow \infty} k^2 D(k^2) = 0. \label{UVpropbehav}
\end{align}
We can see that the gluon propagator fulfills this condition in Yang-Mills theories in the Landau gauge~\cite{OZ80a, OZ80b}. Indeed, the renormalization group (RG) equation for the propagator and the ultraviolet asymptotic freedom yield the asymptotic form for $|k^2| \rightarrow \infty$
\begin{align}
D(k^2) \sim - \frac{Z_{UV}}{k^2 (\ln |k^2|)^\gamma}, \label{eq:section2_UV_asymptotic}
\end{align}
where $Z_{UV}$ is a positive constant and $\gamma$ is the ratio between the coefficient $\gamma_0$ of the leading term of the anomalous dimension and the first coefficient $\beta_0$ of the beta function. For the gluon propagator of the pure Yang-Mills theory in the Landau gauge, $\gamma$ has the value,
\begin{align}
\gamma_0 = - \frac{1}{16 \pi^2} \frac{13}{6} C_2 (G) &< 0, ~~ \beta_0 = - \frac{1}{16 \pi^2} \frac{11}{3} C_2 (G) < 0, \notag \\
\gamma &= \frac{\gamma_0}{\beta_0} = \frac{13}{22} > 0, \label{eq:gamma_0_and_beta_0}
\end{align}
where $C_2(G)$ is the quadratic Casimir invariant of the gauge group $G$. The condition (\ref{UVpropbehav}) holds due to (\ref{eq:section2_UV_asymptotic}) and (\ref{eq:gamma_0_and_beta_0}).

\begin{figure}[tbp]
 \begin{center}
  \includegraphics[width=70mm]{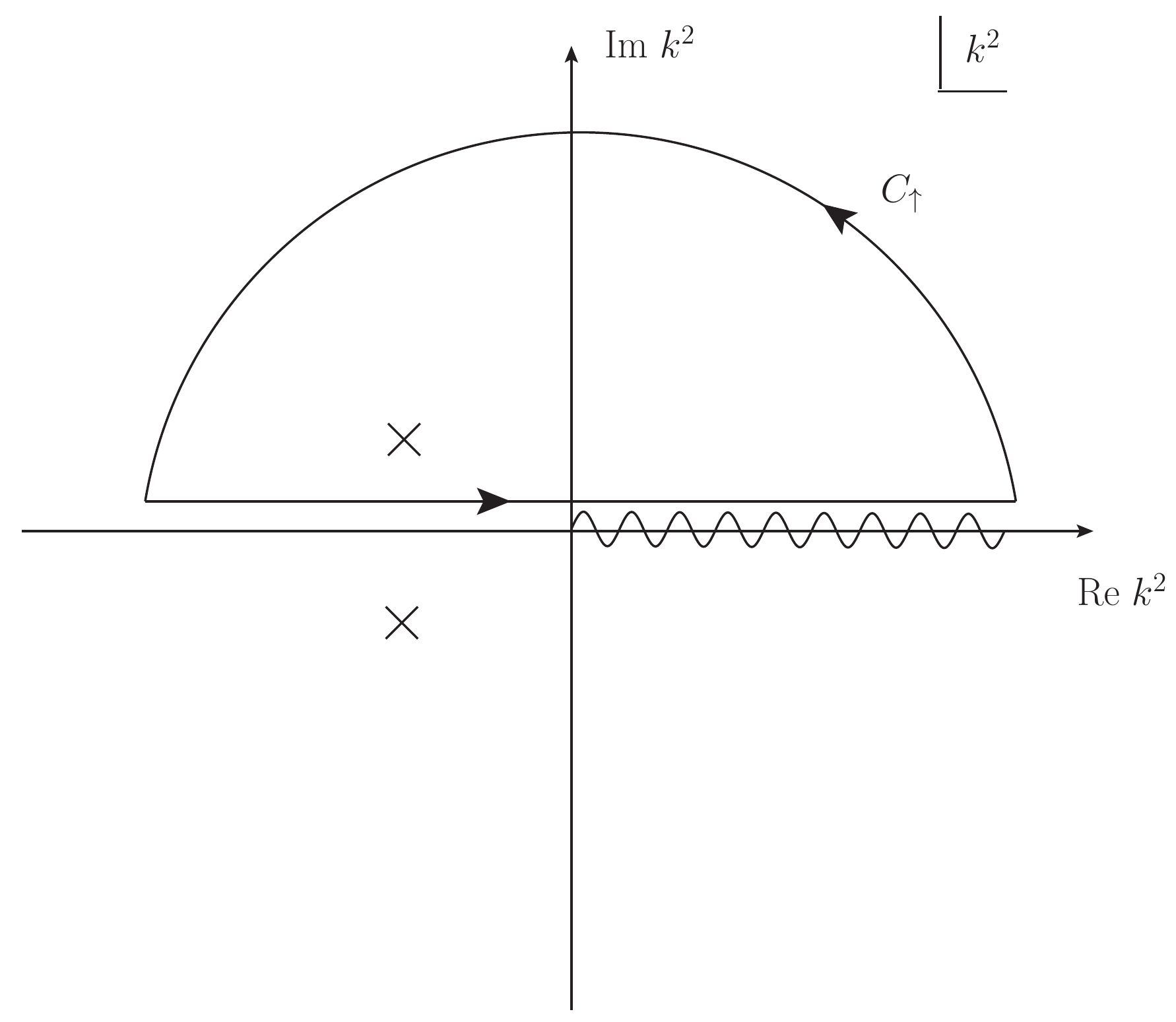}
 \end{center}
 \caption{Contour $C_{\uparrow}$ consisting of the large semicircle and the line $(-\infty + i \epsilon, + \infty + i \epsilon)$ on the complex $k^2$ plane to pick up the pole at $k^2 = v+ i w, \ w > 0$.}
 \label{fig:section2_residue.pdf}
\end{figure}

In this circumstance, the residue $Z$ in (\ref{eq:one_pair_complex}) as a special case of (\ref{eq:spec_repr_complex}) can be measured, according to (\ref{eq:dispersion_residue}),by integrating along the contour $C_{\uparrow}$ consisting of the large semicircle and the line $(-\infty + i \epsilon, + \infty + i \epsilon)$ shown in Fig.~\ref{fig:section2_residue.pdf}. Here, the contour $C_{\uparrow}$ is chosen such that it encloses only the pole located at $k^2 = v + iw $, and the other singularities, namely its complex conjugate pole at $k^2 = v - iw $ and the singularities on the positive real axis, lie outside the contour $C_\uparrow$. We therefore have
\begin{align}
Z &= -\frac{1}{2 \pi i} \oint_{C_{\uparrow}} dk^2 D(k^2).
\end{align}
The assumption (\ref{UVpropbehav}) is enough to eliminate the contribution from  the large semi-circle to obtain
\begin{align}
Z &= -\frac{1}{2 \pi i} \int_{- \infty}^{\infty} dx~ D(x + i \epsilon), ~~ x = {\rm Re}~ k^2. \label{eq:section2_residue_formula}
\end{align}
Substituting the relation according to (\ref{eq:dispersion_complex}),
 \begin{align}
{\rm Im} \ D(x + i \epsilon) = \begin{cases}
0 & \ (x < 0) \\
\pi \rho(x) & \ (x > 0) 
\end{cases}
,
\end{align}
into the real part of (\ref{eq:section2_residue_formula}),
we obtain a novel sum rule for the spectral function,
\footnote{
The imaginary part leads to the relation,
\begin{align}
{\rm Im}~ Z &= \frac{1}{2 \pi} \int_{- \infty}^{\infty} dx~ {\rm Re}~ D(x + i \epsilon). \notag
\end{align}
}
 \begin{align}
2\ {\rm Re} \ Z + \int_0 ^\infty d \sigma^2 \rho (\sigma^2) = 0, \label{eq:modified_superconvergence}
\end{align}
as a consequence of the assumption (\ref{UVpropbehav}) for a propagator of the form (\ref{eq:one_pair_complex}).
Therefore, the usual superconvergence relation \cite{OZ80a, OZ80b},
 \begin{align}
\int_0 ^\infty d \sigma^2 \rho (\sigma^2) = 0, \label{eq:the_superconvergence}
\end{align}
does not hold unless the residue of the complex pole is pure imaginary ${\rm Re} \ Z = 0$ for consistency with the asymptotic freedom and the negativity of the anomalous dimension (\ref{eq:gamma_0_and_beta_0}).

\section{Relation between Unphysical Poles and Winding Number}
In this section, we consider the number of complex poles in a propagator. It can be measured by utilizing the principle of the argument, which is the central claim. The {\it winding number} $N_W(C)$ of the phase of the propagator $D(k^2)$ for a contour $C$ is equal to the difference between the number of zeros $N_Z(\mathcal{D})$ and the number of poles $N_P(\mathcal{D})$ in the region $\mathcal{D}$ bounded by $C$,
\begin{align}
N_W(C) :&= \frac{1}{2 \pi i} \oint_{C} d k^2 \frac{D'(k^2)}{D(k^2)} \notag \\
&= N_Z(\mathcal{D}) - N_P(\mathcal{D}). \label{eq:winding_number_relation}
\end{align}
Here the phase $\theta(k^2)$ is defined by the argument of $D(k^2)$, $D(k^2) = |D(k^2)| e^{i \theta(k^2)}$. Notice that $N_W(C)$ can be expressed as
\begin{align}
N_W(C) = \oint_{C} \frac{d \theta(k^2)}{2 \pi}.
\end{align}
We call poles which are not located on the positive real axis {\it unphysical poles}, and particularly poles located on the negative real axis {\it tachyonic poles}. Therefore, this equation for the contour $C$ presented in Fig.~\ref{fig:section2_analiticity.pdf} provides a relation between the number of unphysical poles and the winding number of a propagator. 

Here we consider the following cases: (A) positive spectral function, (B) negative spectral function, (C) negative spectral function with a massless pole and (D) generalization of these three cases. Further generalizations that could include the scaling solution are presented in Appendix A.
\subsection{Positive spectral function}
(Case I) Suppose that a propagator exhibits the following behaviors.
 \begin{enumerate}
 \item The propagator has the leading asymptotic behavior: $D(z) \sim -\frac{1}{z} \tilde{D}(z)$ as $|z| \rightarrow \infty$, where $\tilde{D}(z)$ is a real and positive function $\tilde{D}(z)>0$ for large $|z|$.
 \item $\rho(\sigma^2) > 0$, i.e., ${\rm Im}\ D(\sigma^2+i\epsilon) > 0$ for $\sigma^2 > 0$.
 \item $D(-\epsilon) > 0$ for sufficiently small $\epsilon > 0$.
 \end{enumerate}
 Let us add some remarks on these assumptions before calculating the winding number.
 First of all, $\tilde{D}(z)$ often depends only on $|z|$: $\tilde{D}(z) = \tilde{D}(|z|)$. For example, $\tilde{D}(z)$ is a constant for the ``physical propagator'' that has the spectral representation, 
 \begin{align}
D_{phys}(k^2) &= \frac{Z}{M^2 - k^2} + \int_{\sigma_0^2} ^\infty d \sigma^2 \frac{\rho(\sigma^2)}{\sigma^2 - k^2}, \notag \\
& Z > 0, \ \ \rho (\sigma^2) > 0.\label{eq:the_physical_propagator}
\end{align}
Indeed, for sufficiently large $|z|$, the physical propagator behaves as
 \begin{align}
D_{phys}(z) \sim - \frac{1}{z} \left[ Z + \int_{\sigma_0^2} ^\infty d \sigma^2 \rho(\sigma^2) \right] = - \frac{1}{z} \tilde{D}(z),
\end{align}
 which shows $\tilde{D}(z)$ is a constant. In the Yang-Mills theory, the RG analysis and the asymptotic freedom give
 \begin{align}
D(z) \sim - \frac{Z_{UV}}{z (\ln |z|)^\gamma}
\end{align}
 as $|z| \rightarrow \infty$. This yields $\tilde{D}(z) = \frac{Z_{UV}}{(\ln |z|)^\gamma} > 0$, which depends only on $|z|$.

 Then, the second assumption ${\rm Im}\ D(\sigma^2+i\epsilon) > 0$ should be understood with the infinitesimal $\epsilon > 0$. By adopting this convention, this case includes the physical propagator of (\ref{eq:the_physical_propagator}). Indeed, we find for any $k^2$
 \begin{align}
{\rm Im}\ D_{phys}(k^2+i\epsilon) &= Z \frac{\epsilon}{(M^2 - k^2)^2 + \epsilon^2} \notag \\
&+ \int_{\sigma_0^2} ^\infty d \sigma^2 \rho(\sigma^2) \frac{\epsilon}{(\sigma^2 - k^2)^2 + \epsilon^2} > 0. \label{convention_of_sign_physical}
\end{align}
 Finally, we take the third assumption to allow the propagator to have a massless pole of the behavior: $D(z) \sim - \frac{Z_0}{z}, \ Z_0 >0$ as $|z| \rightarrow 0$.

\begin{figure}[tbp]
 \begin{center}
 \includegraphics[width=70mm]{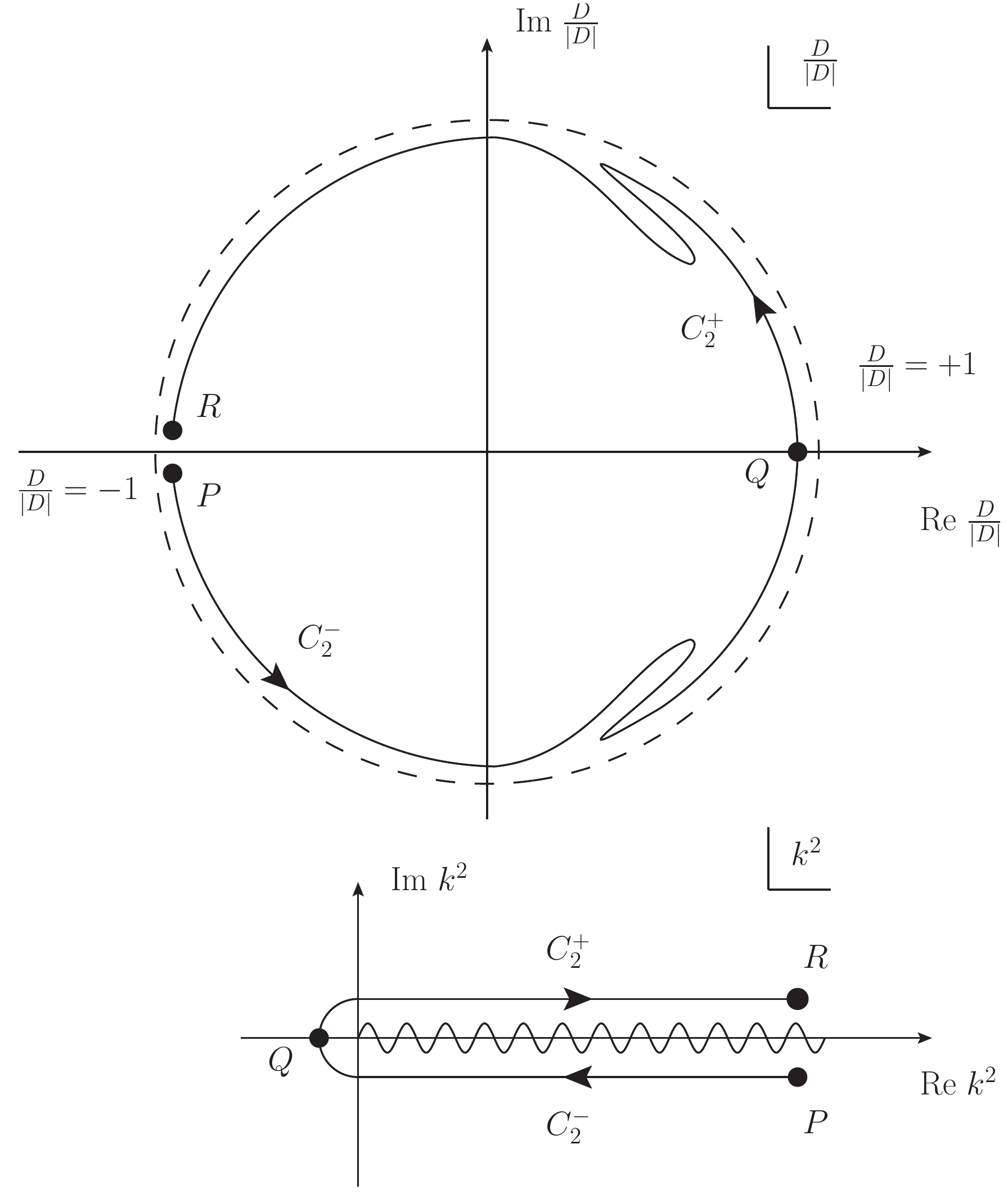}
 \end{center}
 \caption{Schematic picture for the trajectory of the phase factor $e^{i \theta(k^2)} = D(k^2)/|D(k^2)|$ in the case I along $C_2$, consisting of $C_2^-$ (the lower contour from P to Q) and $C_2^+$ (the upper contour from Q to R). The phase factor $D(k^2)/|D(k^2)|$ may change its direction along the lower (resp.~upper) path of $C_2$, however, it does not affect the winding number because of the constraint ${\rm Im}\ D(\sigma^2-i\epsilon) < 0$ (resp.~${\rm Im}\ D(\sigma^2+i\epsilon) > 0$)}
 \label{fig:schematic_c2}
\end{figure}

 The contour $C$ is divided into two paths $C = C_1 + C_2$ as shown in Fig.~\ref{fig:section2_analiticity.pdf}. For the large circle $C_1$, the first assumption (i) leads to
 \begin{align}
N_W (C_1) = \frac{1}{2 \pi i} \int_{C_1} dk^2 \left( - \frac{1}{k^2} \right) = -1.
\end{align}
For the path $C_2$, on the other hand, the phase $\theta(k^2)$ of the propagator $D(k^2)$ rotates in the positive direction to give winding number one as follows. See Fig.~\ref{fig:schematic_c2}. First, the propagator has the phase $\theta = -\pi$ at P: $k^2 = + \infty - i \epsilon$ from the first assumption (i) $D(k^2\rightarrow +\infty) \rightarrow - 0$. Then, the propagator gradually changes its phase along the lower line $C_2^-$ of $C_2$ and reaches $\theta = 0$ at Q: $k^2 = -\epsilon$ from the third assumption (iii) $D(-\epsilon) > 0$ for small $\epsilon > 0$. It is important to note ${\rm Im}\ D(\sigma^2-i\epsilon) < 0$ along the lower line $C_2^-$ from the second assumption (ii). Therefore, the phase $\theta$ moves the half of the phase $S^1$ counterclockwise in total along the lower path $C_2^-$. Similarly, the phase $\theta$ moves the other half of the phase $S^1$ counterclockwise in total along the upper path $C_2^+$ of $C_2$ from Q: $k^2 = - \epsilon$ to R: $k^2 = +\infty +i\epsilon$. The contribution of $C_2$ becomes
 \begin{align}
N_W(C_2) = +1.
\end{align}
Hence, the net winding number is zero $N_W(C) = N_W(C_1) + N_W(C_2) = 0$, and we obtain
 \begin{align}
N_P = N_Z.
\end{align}
It is consistent with the fact that the physical propagator (\ref{eq:the_physical_propagator}) has no unphysical poles $N_P = 0$ if it has no zeros $N_Z = 0$.

\subsection{Negative spectral function}
(Case II) We investigate a propagator with the following properties
 \begin{enumerate}
 \item The propagator has the leading asymptotic behavior: $D(z) \sim -\frac{1}{z} \tilde{D}(z)$ as $|z| \rightarrow \infty$, where $\tilde{D}(z)$ is a real and positive function for large $|z|$.
 \item $\rho(\sigma^2) < 0$, i.e., ${\rm Im}\ D(\sigma^2+i\epsilon) < 0$ for $\sigma^2 > 0$.
 \item $D(k^2 = 0) > 0$.
 \end{enumerate}
 They are in fact properties of the gluon propagator in the one-loop massive Yang-Mills model as we will see in Section V. The crucial difference from the case I of a positive spectral function is the second assumption (ii) of the negative spectral function.
 
For the large circle $C_1$, we have
 \begin{align}
N_W(C_1) = -1.
\end{align}
However, the second assumption (ii) ${\rm Im}\ D(\sigma^2+i\epsilon) < 0$ and ${\rm Im}\ D(\sigma^2-i\epsilon) > 0$ shows that the phase of the propagator rotates along $C_2$ in the negative direction with winding number one. Therefore, the contribution of $C_2$ becomes
 \begin{align}
N_W(C_2) = -1.
\end{align}
We hence obtain $N_W(C) = -2$ and
 \begin{align}
N_P = N_Z + 2. \label{eq:case_2_number_of_complex_poles}
\end{align}
This shows the existence of $N_Z+2$ unphysical poles. In particular, if the propagator has no zeros $N_Z = 0$, the number of unphysical poles is two $N_P =2$. This implies that the propagator has a pair of complex conjugate poles or tachyonic poles, poles on the negative real axis, with multiplicity two.

\subsection{Negative spectral function and massless pole}
\begin{figure}[tbp]
\begin{center}
 \includegraphics[width=70mm]{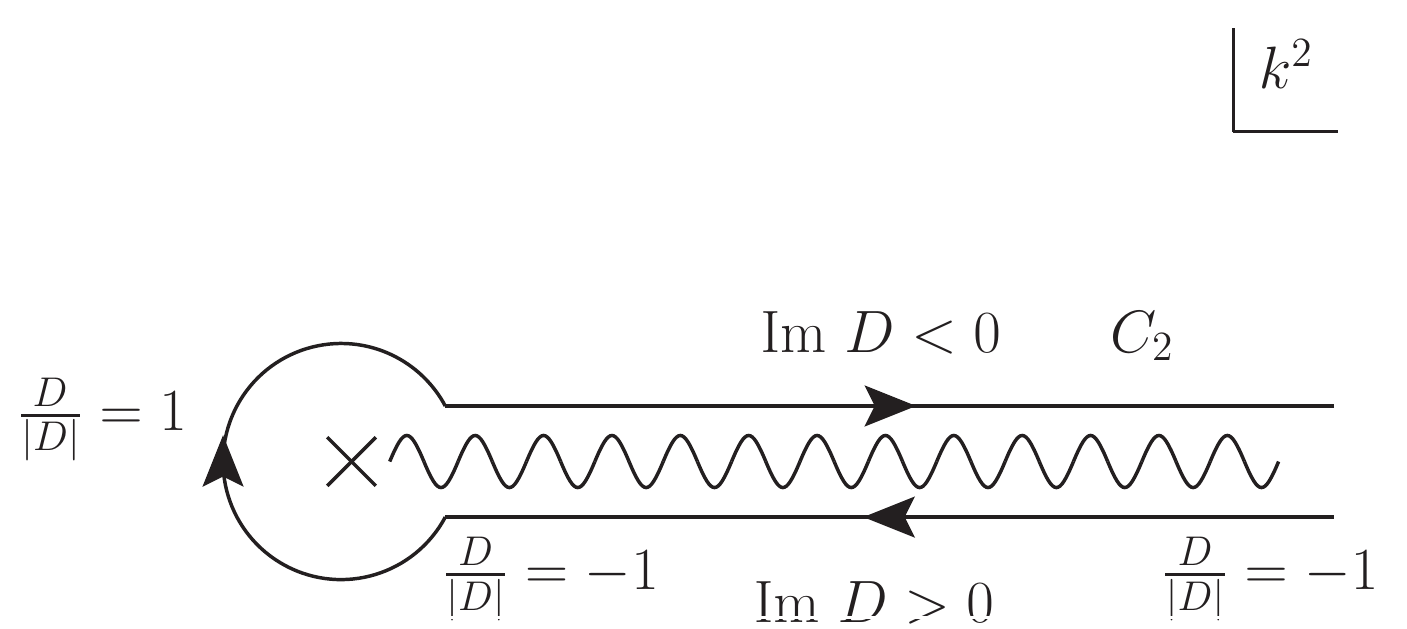}
\end{center}
 \caption{ The contribution of the $C_2$ integral in the case of $Z_c > 0$. This shows the winding number is zero on the real axis, while the winding number is $+1$ around the massless pole with $Z_c > 0$.}
 \label{fig:ghost_complex_plane.pdf}
\end{figure}

(Case III) We consider a propagator satisfying the following properties.
 \begin{enumerate}
 \item The propagator has the leading asymptotic behavior: $D(z) \sim -\frac{1}{z} \tilde{D}(z)$ as $|z| \rightarrow \infty$, where $\tilde{D}(z)$ is a real and positive function for large $|z|$.
 \item $\rho(\sigma^2) < 0$, i.e., ${\rm Im}\ D(\sigma^2+i\epsilon) < 0$ for $\sigma^2 > 0$.
 \item $D(z) \sim - Z_c/z$ with real $Z_c$ as $|z| \rightarrow 0$.
 \end{enumerate}
 They are in fact properties of the ghost propagator in the one-loop massive Yang-Mills model, up to an overall sign, as we will see in Sec.~V. The evaluation of the integral along $C_1$ is the same:
\begin{align}
N_W(C_1) = -1.
\end{align}
However, more careful analysis is required for the $C_2$ integral in this case. If $Z_c < 0$, the winding number of the phase along $C_2$ is zero, $N_W(C_2) = 0$, because ${\rm Im}\ D(\sigma^2+i\epsilon) < 0$ and ${\rm Im}\ D(\sigma^2-i\epsilon) > 0$ holds for $-\delta < \sigma^2$ with a sufficiently small $\delta > 0$ and the propagator is negative both at $k^2 = + \infty$ and at $k^2 = -\epsilon$. For the case of $Z_c > 0$, the winding number of the phase along $C_2$ is one, $N_W(C_2) = 1$, since only the small circle contributes to the integral. See Fig.~\ref{fig:ghost_complex_plane.pdf}.
 Therefore, we obtain the result
\begin{align}
N_P = 
\begin{cases}
N_Z \ \ &(Z_c > 0) \\
N_Z+1 \ \ &(Z_c < 0).
\end{cases}
\label{eq:case_3_complex_poles}
\end{align}
If the propagator has no zeros, the propagator has a tachyonic pole if $Z_c < 0$ and has no unphysical pole if $Z_c > 0$, besides the pole at the origin $k^2 = 0$.

\subsection{Quasipositive or quasinegative spectral function}
We discuss generalizations of the cases examined above. For this purpose, let us define some properties of spectral functions. We call a spectral function {\it quasipositive} (resp.~{\it quasinegative}) if and only if $\rho(k_0^2) > 0$ (resp.~$\rho(k_0^2) < 0$) at all real and positive zeros $k_0^2$ of ${\rm Re}\ D(k^2)$ {\it i.e.}, ${\rm Re}\ D(k_0^2) = 0$ ($k_0^2 > 0$). For example, a negative (resp.~positive) spectral function is always quasinegative (resp.~quasipositive). 

 \begin{figure}[tbp]
\begin{center}
 \includegraphics[width=70mm]{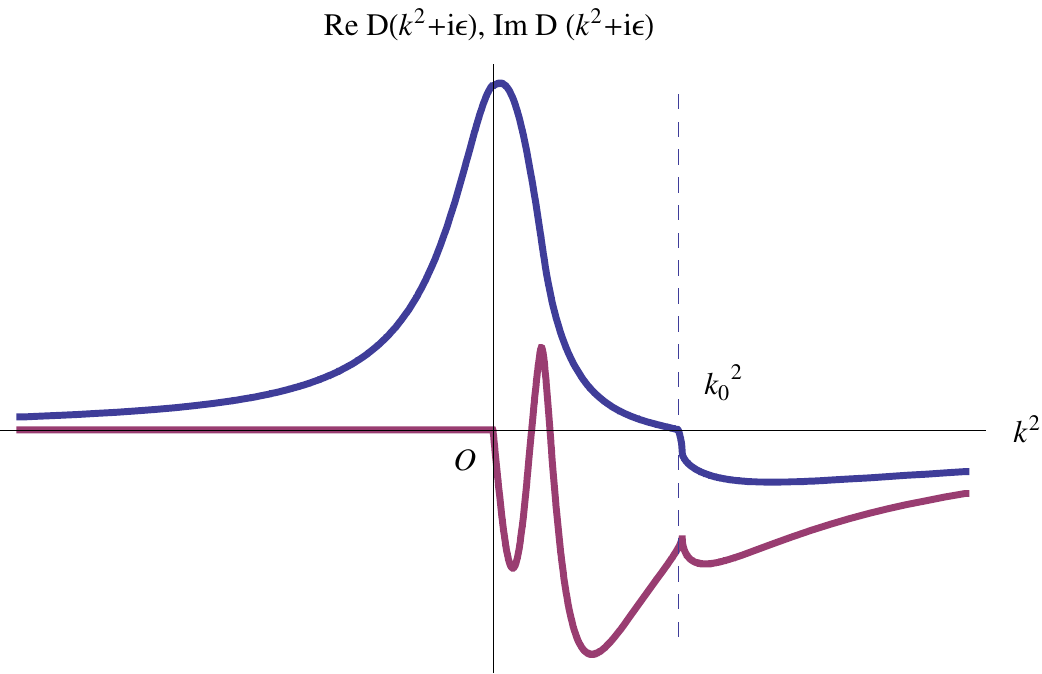}
\end{center}
 \caption{Schematic example of a quasinegative spectral function. The blue line plots ${\rm Re}\ D(k^2+i\epsilon)$ and the red line plots ${\rm Im}\ D(k^2+i\epsilon) = \pi \rho(k^2) $.  The dashed vertical line shows the location of the zero $k_0 ^2$ of ${\rm Re}\ D(k^2+i\epsilon)$, at which $\rho(k^2)$ takes a negative value.}
 \label{fig:Siringoexample}
\end{figure}

Note that if ${\rm Re}\ D(k^2)$ has one zero $k_0^2 > 0$, any spectral function can be classified into quasipositive or quasinegative depending on the sign of $\rho(k_0^2)$. The case $\rho(k_0^2) = 0$ is excluded as the sign of $\rho(k^2_0) = \frac{1}{\pi} {\rm Im}\ D(\sigma^2+i\epsilon) $ is understood with the infinitesimal $\epsilon>0$ as mentioned around (\ref{convention_of_sign_physical}). Indeed, $\rho(k_0^2)$ is nonzero in this sense because the propagator $D(z)$ can have only isolated zeros due to the holomorphicity. Figure~\ref{fig:Siringoexample} illustrates an example of a quasinegative spectral function.

We then give generalizations of cases I, II, and III.

\begin{figure}[tbp]
 \begin{center}
 \includegraphics[width=70mm]{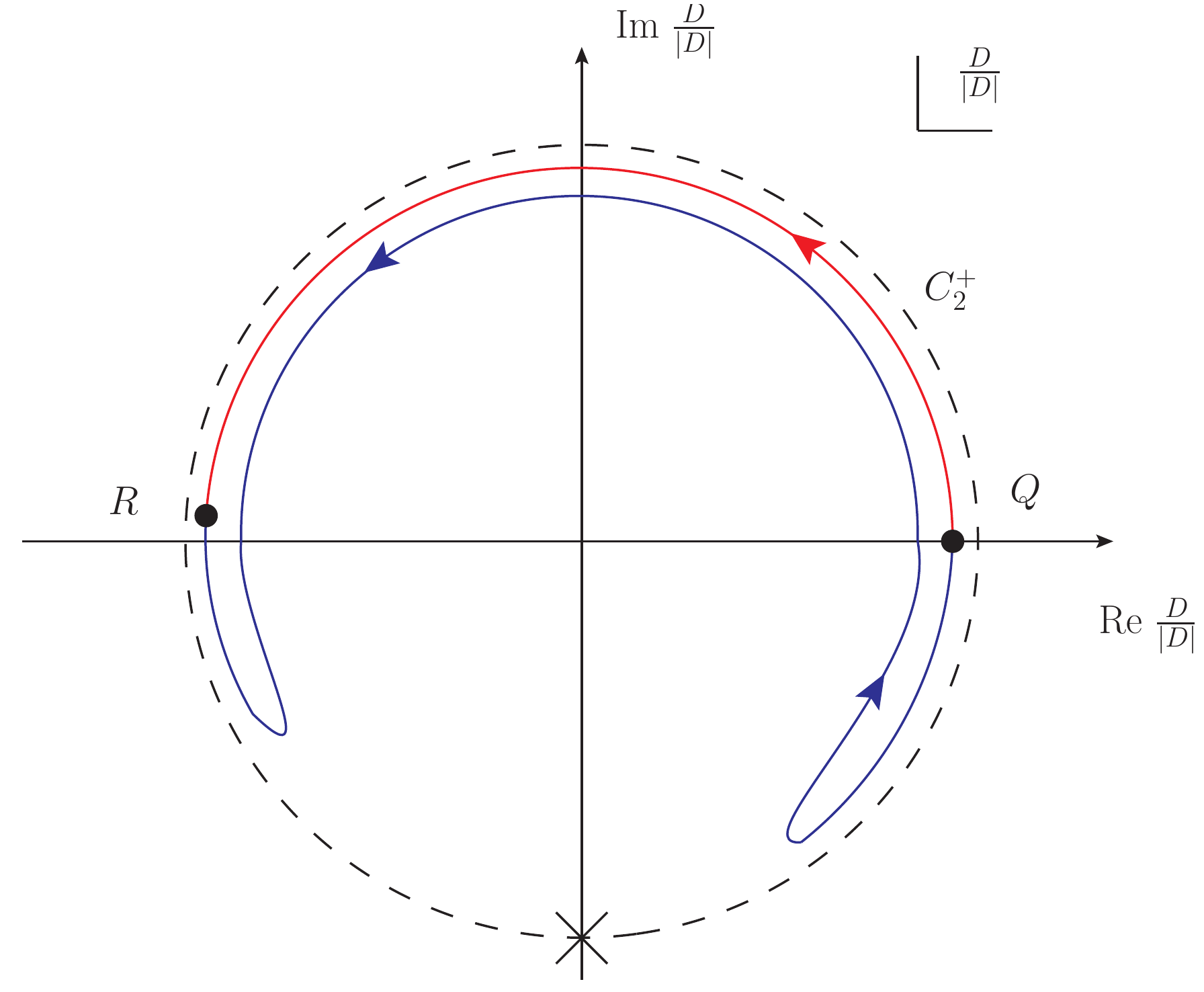}
 \end{center}
 \caption{Schematic picture of phase trajectories along $C_2^+$, the upper contour of $C_2$ from Q to R depicted in Fig.~\ref{fig:schematic_c2}. The blue curve shows the phase trajectory with a quasipositive spectral function, which does not pass through the point $D/|D| = - i$. We can continuously deform the blue trajectory into the red one with a positive spectral function.}
 \label{fig:schematic_continuous_deformation}
\end{figure}

(Case I') Suppose that a propagator $D(z)$ exhibits the following properties: 
 \begin{enumerate}
 \item The propagator has the leading asymptotic behavior: $D(z) \sim -\frac{1}{z} \tilde{D}(z)$ as $|z| \rightarrow \infty$, where $\tilde{D}(z)$ is a real and positive function for large $|z|$.
 \renewcommand{\labelenumi}{(\roman{enumi}')}
 \item The spectral function $\rho(\sigma^2)$ is quasipositive.
 \renewcommand{\labelenumi}{(\roman{enumi})}
 \item $D(-\epsilon) > 0$ for small $\epsilon > 0$.
 \end{enumerate}
 Then we obtain
\begin{align}
N_P= N_Z. 
\end{align}
This result is shown by a similar evaluation based on the invariance of the winding number under continuous deformations. Indeed, if the spectral function is quasipositive, the phase trajectory $D/|D|$ (see Fig.~\ref{fig:schematic_c2}) along the lower (resp.~upper) contour $C_2^-$ (resp.~$C_2^+$) of $C_2$ does not pass through the point $D/|D| = + i$ (resp.~$D/|D| = - i$). Thus, the phase trajectory of the case I' can be continuously deformed into that of the case I with a positive spectral function. For example, a schematic picture of the continuous deformation for the contour $C_2^+$ is displayed in Fig.~\ref{fig:schematic_continuous_deformation}.

The similar arguments can be applied to the cases II and III. Therefore, we have the following propositions.

(Case II') Suppose that a propagator $D(z)$ exhibits the following properties: 
 \begin{enumerate}
 \item The propagator has the leading asymptotic behavior: $D(z) \sim -\frac{1}{z} \tilde{D}(z)$ as $|z| \rightarrow \infty$, where $\tilde{D}(z)$ is a real and positive function for large $|z|$.
\renewcommand{\labelenumi}{(\roman{enumi}')}
 \item The spectral function $\rho(\sigma^2)$ is quasinegative.
 \renewcommand{\labelenumi}{(\roman{enumi})}
 \item $D(k^2 = 0) > 0$.
 \end{enumerate}
Then we have
 \begin{align}
N_P = N_Z + 2,
\end{align}
by the invariance of the winding number under continuous deformations.

(Case III') Suppose that a propagator $D(z)$ exhibits the following properties: 
 \begin{enumerate}
 \item The propagator has the leading asymptotic behavior: $D(z) \sim -\frac{1}{z} \tilde{D}(z)$ as $|z| \rightarrow \infty$, where $\tilde{D}(z)$ is a real and positive function for large $|z|$.
\renewcommand{\labelenumi}{(\roman{enumi}')}
 \item The spectral function $\rho(\sigma^2)$ is quasipositive or quasinegative for $\sigma^2 > 0$.
 \renewcommand{\labelenumi}{(\roman{enumi})}
 \item $D(z) \sim - Z_c/z$ with real $Z_c$ as $|z| \rightarrow 0$.
 \end{enumerate}
 Then we obtain
\begin{align}
N_P = 
\begin{cases}
N_Z \ \ &(Z_c > 0) \\
N_Z+1 \ \ &(Z_c < 0, \ \rho\mbox{ is quasinegative}) \\
N_Z-1 \ \ &(Z_c < 0, \ \rho\mbox{ is quasipositive}). \label{eq:case3prime}
\end{cases}
\end{align}
This claim is shown as follows. As Fig.~\ref{fig:ghost_complex_plane.pdf} shows, if the spectral function is quasipositive or quasinegative, the integral on the positive real axis $(+ \epsilon, + \infty)$ does not affect the winding number for $Z_c > 0$, which yields $N_W(C_2) = 1$. If $Z_c < 0$, the upper and lower paths on the positive real axis $(+ \epsilon, + \infty)$ give a nontrivial winding number, which is identified with that of a positive or negative spectral function due to the invariance under continuous deformations. Therefore, $N_W(C_2) = 0$ for a quasinegative spectral function as before, and $N_W(C_2) = 2$ for a quasipositive spectral function. They lead to the result (\ref{eq:case3prime}).

\section{MASSIVE YANG-MILLS MODEL AT ONE-LOOP AS AN EFFECTIVE MODEL}
As a preliminary remark, the massive Yang-Mills model is an effective model of the Landau-gauge Yang-Mills theory, which reproduces the numerical lattice results of the gluon and ghost propagators well. Although it is possible to discuss the massive Yang-Mills model with a more general gauge, it does not motivate us since we do not have evidence that it is an effective model of the Yang-Mills theory in the other gauge, or related to some other models. In this section, we review the results on one-loop quantum corrections for the massive Yang-Mills model. The Lagrangian of the model is given by
\begin{align}
{\mathscr L}_{mYM} &= {\mathscr L}_{YM} + {\mathscr L}_{GF} + {\mathscr L}_{FP} + {\mathscr L}_{m}, \\
{\mathscr L}_{YM} &= - \frac{1}{4} {\mathscr F}^A_{\mu \nu} {\mathscr F}^{A \mu \nu}, \notag \\
{\mathscr L}_{GF} &= - \frac{1}{2 \alpha_B} (\partial^\mu {\mathscr A}^A_\mu )^2 \notag \\
{\mathscr L}_{FP} &= i \bar{{\mathscr C}}^A \partial^\mu {\mathscr D}_\mu[{\mathscr A}]^{AB} {\mathscr C}^B \notag \notag \\
&= i \bar{{\mathscr C}}^A \partial^\mu (\partial_\mu {\mathscr C}^A + g_B f^{ABC} {\mathscr A}^B {\mathscr C}^C) \notag \\
{\mathscr L}_{m} &= \frac{1}{2} M^2_B {\mathscr A}_\mu ^A {\mathscr A}^{A \mu}. \notag 
\end{align}
We obtain the Landau gauge results by taking the limit $\alpha_B \rightarrow 0$ after calculating with arbitrary $\alpha_B > 0$ \cite{Suda18}. We introduce the renormalization factors $(Z_A,Z_C, Z_{\bar{C}} = Z_C),\ Z_g,\ Z_{M^2}$ for the gluon, ghost and antighost fields $({\mathscr A}_\mu, {\mathscr C}, \bar{{\mathscr C}})$, the gauge coupling constant $g$, and the mass parameter $M^2$ respectively:
\begin{align}
{\mathscr A}^\mu &= \sqrt{Z_A} {\mathscr A}_R^\mu, \ \ {\mathscr C} = \sqrt{Z_C} {\mathscr C}_R, \ \ \bar{{\mathscr C}} = \sqrt{Z_C} \bar{{\mathscr C}}_R, \notag \\
g_B &= Z_g g, \ \ M^2_B = Z_{M^2} M^2.
\end{align}
%%%% The origin of the phenomenological mass term can be connected with the Gribov problem[Serreu--Tissier][Tissier2017], some condensates[H. Verschelde, K. Knecht, K. Van Acoleyen, and M. Vanderkelen,2001] or other non-perturbative effects. It can be, moreover, related with the gauge-scalar model up to Gribov copies. [Kondo2018].

For comparison with the lattice data, we work with Euclidean signature to describe the one-loop results. We will subsequently perform the analytic continuation of them to the complex $k^2$ momentum plane. For the gluons, we introduce the two-point vertex function $\Gamma_{{\mathscr A}}^{(2)}$, the transverse propagator ${\mathscr D}_T$, and the vacuum polarization $\Pi_T$ together with their dimensionless ones denoted by $\hat{\Gamma}, \ \hat{\Pi}$ as
\begin{align}
\Gamma_{{\mathscr A}}^{(2)}(k_E) &:= [{\mathscr D}_T (k_E)]^{-1} \notag \\
&= k^2_E + M^2 + \Pi_T (k_E) + k_E^2 \delta_Z + M^2 \delta_{M^2} \notag \\
&=: M^2 \hat{\Gamma}(s) =: M^2 [s+1 + \hat{\Pi}(s) + s \delta_Z + \delta_{M^2}] \notag \\
&=: M^2 [s+1 + \hat{\Pi}^{ren}(s)], 
\end{align}
where $k_E$ denotes the Euclidean momentum ($k^2 = - k_E^2$), $s := k_E^2/M^2$, and $\delta_Z = Z_A - 1$, $\delta_{M^2} = Z_A Z_{M^2} -1$ are the counterterms, which are determined by imposing renormalization conditions. Similarly, we introduce the ghost counterparts for the two-point vertex function $\Gamma_{gh}^{(2)}$, the propagator $\Delta_{gh}$ and the self-energy $ \Pi_{gh}$ together with their dimensionless versions $\hat{\Gamma}_{gh},\ \hat{\Pi}_{gh}$ as
\begin{align}
\Gamma_{gh}^{(2)}(k_E) &:= - [\Delta_{gh} (k_E)]^{-1} = k^2_E + \Pi_{gh} (k_E) + k_E^2 \delta_C \notag \\
&=: M^2 \hat{\Gamma}_{gh}(s) =: M^2 [s + \hat{\Pi}_{gh}(s) + s \delta_C] \notag \\
&=: M^2 [s + \hat{\Pi}_{gh}^{ren}(s) ], \label{eq:ghost_def}
\end{align}
where $\delta_C = Z_C - 1$ is the counterterm. The minus sign has been introduced since the ghost state has a negative norm at the tree level.

The results of the one-loop calculations by the dimensional regularization are as follows \cite{Suda18, TW10, TW11}. The vacuum polarization for gluons reads
\begin{align}
\hat{\Pi}(s) &= \frac{g^2 C_2 (G)}{192 \pi^2} s \Biggl[ \left( \frac{9}{s} - 26 \right) \left[ \varepsilon^{-1} - \gamma + \ln (4 \pi) + \ln M^{-2} \right], \notag \\
&- \frac{121}{3} + \frac{63}{s} + h(s) \Biggr] \label{eq:dimregvacpol} \\
h(s) &:= - \frac{1}{s^2} + \left( 1- \frac{s^2}{2} \right) \ln s \notag \\
&+ \left( 1+ \frac{1}{s}\right)^3 (s^2 - 10s + 1) \ln (s+1) \notag \\
&+ \frac{1}{2} \left( 1+ \frac{4}{s} \right)^{3/2} (s^2 - 20 s + 12) \ln \left( \frac{\sqrt{4+s} - \sqrt{s}}{\sqrt{4+s} + \sqrt{s}} \right). \label{eq:the_function_h_of_s}
\end{align}
where $\varepsilon := 2 - D/2$, $\gamma$ is the Euler-Mascheroni constant, and $C_2(G)$ is the quadratic Casimir invariant of the group $G$. For the ghosts, we obtain
\begin{align}
\hat{\Pi}_{gh}(s) &= \frac{g^2 C_2 (G)}{64 \pi^2} s \biggl[ -3 \left[ \varepsilon^{-1} - \gamma + \ln (4 \pi) + \ln M^{-2} \right] \notag \\
& -5 + f(s) \biggr], \\
f(s) &:= - \frac{1}{s} - s \ln s + \frac{(1+s)^3}{s^2} \ln (s+1) .
\end{align}
The divergent terms $\varepsilon^{-1} - \gamma + \ln (4 \pi) $ can be eliminated by the counterterms. Notice that the divergent part of $\delta_C + \delta_{M^2}$ is zero, which reflects the nonrenormalization theorem \cite{DVS03}
\begin{align}
Z_A Z_C Z_{M^2} = 1.
\end{align}
 We can take, for instance, the following renormalization conditions adopted by Tissier and Wschebor \cite{TW11}: 
\begin{align}
{\rm [TW]}
 \begin{cases}
 Z_A Z_C Z_{M^2} = 1 \\
 \Gamma_{{\mathscr A}}^{(2)} (k_E = \mu) = \mu^2 + M^2\\
 \Gamma_{gh}^{(2)}(k_E = \mu) = \mu^2
 \end{cases}
 \Leftrightarrow \ 
 \begin{cases}
 \delta_C + \delta_{M^2} = 0 \\
 \hat{\Pi}^{ren}(s = \nu) = 0 \\
 \hat{\Pi}_{gh}^{ren} (s = \nu) = 0, \label{TWrenormalization_condition}
 \end{cases}
\end{align}
where $\nu := \mu^2/M^2$ in the one-loop level and Taylor scheme \cite{Taylor71}
\begin{align}
Z_g \sqrt{Z_A} Z_C = 1,
\end{align}
for the coupling renormalization. The renormalized vacuum polarizations $\hat{\Pi}^{TW}_{ren.} (s)$ and $\hat{\Pi}^{TW}_{gh,ren.} (s)$ under this renormalization conditions are,
\begin{align}
\hat{\Pi}^{TW}_{ren.}(s) &= \frac{g^2 C_2 (G)}{192 \pi^2} s \Biggl[ \frac{48}{s} + h(s) + \frac{3 f(\nu)}{s} - (s \rightarrow \nu) \Biggr], \label{TWgluonvacuumpolarization} \\
\hat{\Pi}_{gh,ren.}^{TW}(s) &= \frac{g^2 C_2 (G)}{64 \pi^2} s \biggl[ 
 f(s) - f(\nu) \biggr]. \label{TWghostvacuumpolarization}
\end{align}
Note that the propagator satisfies
\begin{align}
{\mathscr D}_T (k_E = 0) = \frac{1}{M^2 [1 + \hat{\Pi}^{TW}_{ren.} (0)]}> 0. \label{pos_IR_prop_cond}
\end{align}
Indeed, the vacuum polarization at $s=0$ reads
\begin{align}
\hat{\Pi}^{TW}_{ren.}(s=0) &= \frac{g^2 C_2 (G)}{192 \pi^2} s \Biggl[ \frac{48}{s} - \frac{111}{2s} + \frac{3 f(\nu)}{s} \Biggr]\Biggr|_{s=0} \notag \\
&= \frac{g^2 C_2 (G)}{192 \pi^2} \Biggl[ 3 f(\nu) - \frac{15}{2} \Biggr]>0,
\end{align}
where we have used $h(s) = - \frac{111}{2s} + O(\ln s)$ and the fact that $f(s)$ is a monotonically increasing function with $f(0) = 5/2$. This shows ${\mathscr D}_T (k_E = 0) > 0$.

Note that if one chooses the following parameters at $\mu_0 = 1$ GeV,
\begin{align}
g = 4.1 \ \ M =0.45\ {\rm GeV}, \label{eq:physical_value}
\end{align}
the propagators in the strict one-loop level are in good accordance with the results of the numerical lattice simulations of $G = SU(3)$~\cite{DOS16, KSOMH18}.

Note that, by adopting the renormalization condition [TW], there is an ``infrared-safe'' parameter region $(M^2,g^2)$, where the running coupling remains finite without Landau pole~\cite{TW11}. This allows the systematic RG improvement in all energy scales. In the infrared-safe region, the propagators present the decoupling features, and on the separatrix between the infrared-safe and Landau pole region, the propagators exhibit the Gribov-type scaling solution~\cite{RSTW17}. When we discuss RG improvements of this model, we will utilize these facts.

In what follows, we assume that a suitable renormalization condition is adopted to satisfy the positivity of (\ref{pos_IR_prop_cond}). Then, the two-point vertex functions have the following form
\begin{align}
\hat{\Gamma}(s) &= 1 + s + \hat{\Pi}^{TW}_{ren.}(s) + \delta'_Z s + \delta'_{M^2} , \notag \\
& = 1 + s + \hat{\Pi}^{ren}(s), \\
\hat{\Gamma}_{gh}(s) &= s + \hat{\Pi}_{gh,ren.}^{TW}(s) + \delta'_C s \notag \\
&= s + \hat{\Pi}_{gh}^{ren}(s).
\end{align}

\section{ANALYTIC STRUCTURE OF THE PROPAGATORS}
In this section, we examine the analytic structure of the propagators of the massive Yang-Mills model. Using the method described in Sec.~III, we show that the number of unphysical poles is two for the gluon propagator. Similarly, we also show that the ghost propagator has at most one tachyonic pole, {\it i.e.}, a real negative pole, and no complex conjugate pole. Moreover, we investigate the pole structure of the propagator in the parameter space $(M^2,g^2)$ in the renormalization condition [TW], (\ref{TWrenormalization_condition}).
\subsection{Negativity of the gluon spectral function}
\begin{figure}[tbp]
 \begin{center}
 \includegraphics[width=70mm]{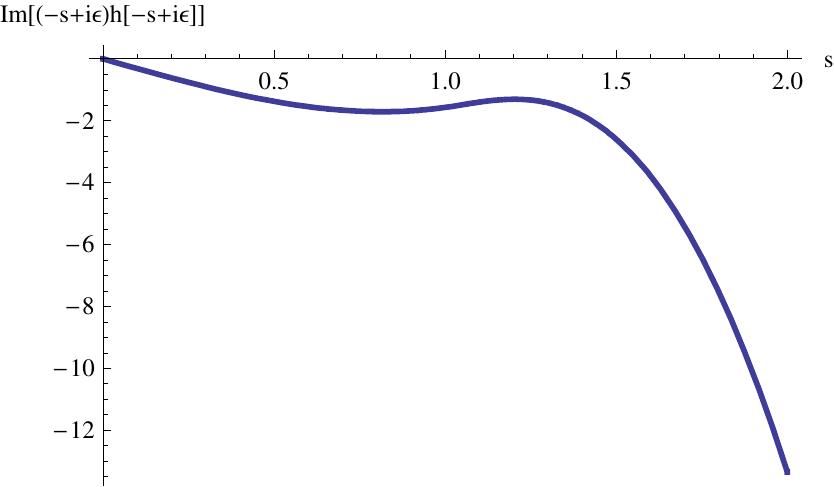}
 \includegraphics[width=70mm]{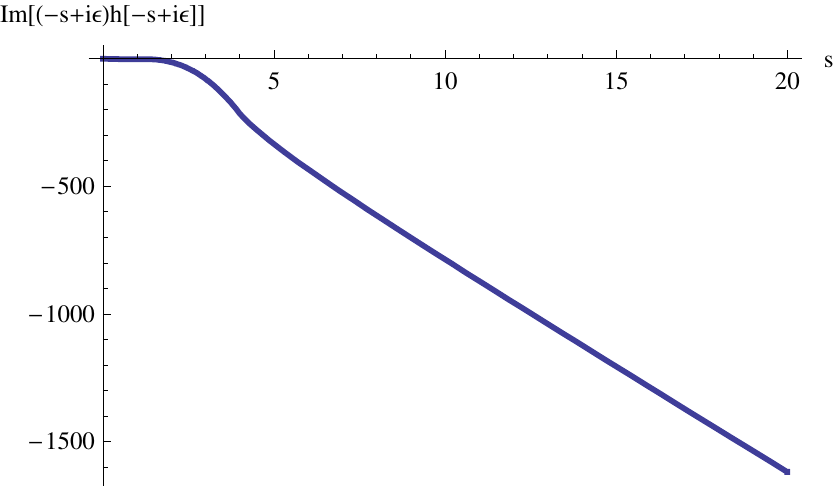}
 \end{center}
 \caption{Plots of ${\rm Im}\ [(- s + i\epsilon) h(-s + i \epsilon) ]$ for different ranges of $s$. They indicate $\rho(\sigma^2) < 0$. }
 \label{fig:ImPi}
\end{figure}

\begin{figure}[tbp]
 \begin{center}
 \includegraphics[width=70mm]{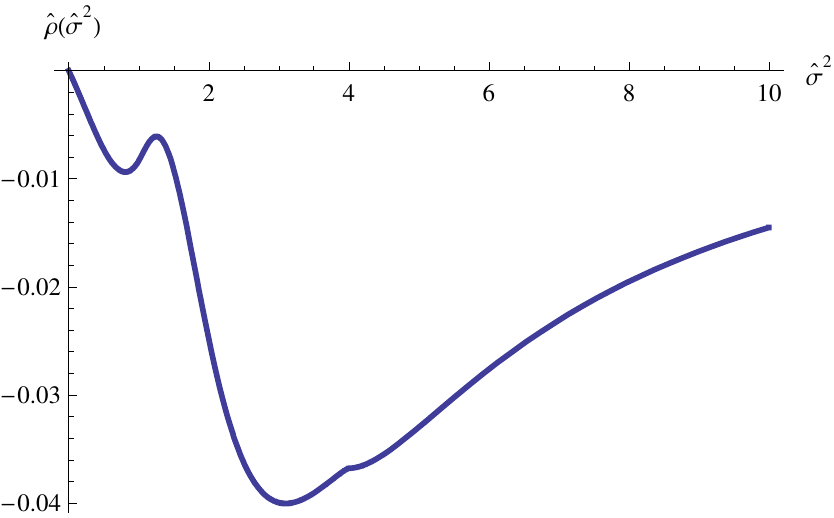}
 \end{center}
 \caption{ Dimensionless spectral function $\hat{\rho}(\hat{\sigma}^2)$ at the physical parameters $g = 4.1, \ M =0.45\ {\rm GeV},$ with [TW] renormalization condition and $G = SU(3)$.}
 \label{fig:spectral_hat_rho_gluon}
\end{figure}

Here we establish the negativity of the gluon spectral function in order to apply the result of section III.
We calculate the spectral function $\rho(\sigma^2)$ from the imaginary part of the propagator in the Minkowski region, obtained by the analytic continuation, according to the common dispersion relation (\ref{eq:dispersion_complex}),
\begin{align}
\rho(\sigma^2) &= \frac{1}{\pi} {\rm Im}\ {\mathscr D}_T (\sigma^2+i\epsilon) \notag \\
 &= \frac{1}{\pi} {\rm Im}\ {\mathscr D}_T (k_E^2 = -\sigma^2-i\epsilon) \notag \\
 &= - \frac{1}{\pi M^2} \left. \frac{ {\rm Im}\ \hat{\Pi}(\tilde{s})}{\left[ 1 + \tilde{s} + {\rm Re}\ \hat{\Pi}(\tilde{s})\right]^2 + \left[ {\rm Im}\ \hat{\Pi} (\tilde{s}) \right]^2} \right|_{\tilde{s} = - \frac{\sigma^2}{M^2} - i \epsilon} \notag \\
 &=: \frac{\hat{\rho}(\sigma^2/M^2)}{M^2}.
\end{align}
Therefore, the negativity of the gluon spectral function $\rho$ follows from the negativity of the imaginary part of the vacuum polarization $\hat{\Pi}$,
\begin{align}
\rho(\sigma^2) < 0 \ &\Leftrightarrow \ - {\rm Im}\ \hat{\Pi}(- \frac{\sigma^2}{M^2} - i \epsilon) < 0 \notag \\
& \Leftrightarrow \ {\rm Im}\ \hat{\Pi}(- \frac{\sigma^2}{M^2} + i \epsilon) < 0 .
\end{align}
Indeed, ${\rm Im}\ \hat{\Pi}(- s + i \epsilon)$ can be evaluated from (\ref{eq:dimregvacpol}) and (\ref{eq:the_function_h_of_s}) as 
\begin{align}
{\rm Im}\ \hat{\Pi}(- s + i \epsilon) = \frac{g^2 C_2 (G)}{192 \pi^2} {\rm Im}\ [(- s + i\epsilon) h(-s + i \epsilon) ] < 0. \label{gluonnegativeity}
\end{align}
Note that ${\rm Im}\ \hat{\Pi}(- s + i \epsilon)$ does not depend upon renormalization conditions, since the counterterms do not contribute to the imaginary part. Moreover, its sign is independent of the parameters $(M^2,g^2)$. In Fig.~\ref{fig:ImPi}, we give plots of ${\rm Im}\ [(- s + i\epsilon) h(-s + i \epsilon) ]$, which guarantees that $\rho(\sigma^2) < 0$ for any $\sigma^2 > 0$ and parameters $M^2, \ g^2 > 0$. The dimensionless spectral function $\hat{\rho} (\hat{\sigma}^2)$ at the values of the parameters (\ref{eq:physical_value}) with the renormalization condition [TW] is plotted in Fig.~\ref{fig:spectral_hat_rho_gluon}

\subsection{Complex poles}
We now show that the number of unphysical poles of the gluon propagator is two. We examine the positions of the complex poles for various parameters $(M^2,g^2)$ in the [TW] renormalization condition.

First, note that ${\rm Im}\ \hat{\Pi}(- s + i \epsilon) \neq 0$ for $s>0$ yields that the gluon propagator ${\mathscr D}_T(k^2)$ has no pole on the real positive axis.
If we assume the analyticity except on the positive real axis, $\rho(\sigma^2) < 0$ for $\sigma^2 > 0$ contradicts the fact that the gluon propagator $D(k_E^2)$ is positive for some $k_E^2$, since the spectral representation (\ref{eq:KL_spectral_repr}) gives
 \begin{align}
D(k_E^2) = \int_0 ^\infty d \sigma^2 \frac{\rho(\sigma^2)}{\sigma^2 + k_E^2}
\end{align}
in the Euclidean momentum $k_E^2$. From the explicit expressions of the propagators given in Sec.~IV, the ghost and gluon propagators have branch cuts only on the positive real axis of the complex $k^2$ plane. Thus, there must exist some poles in the unphysical region excluding the positive real axis.

\begin{figure}[tbp]
 \begin{center}
 \includegraphics[width=70mm]{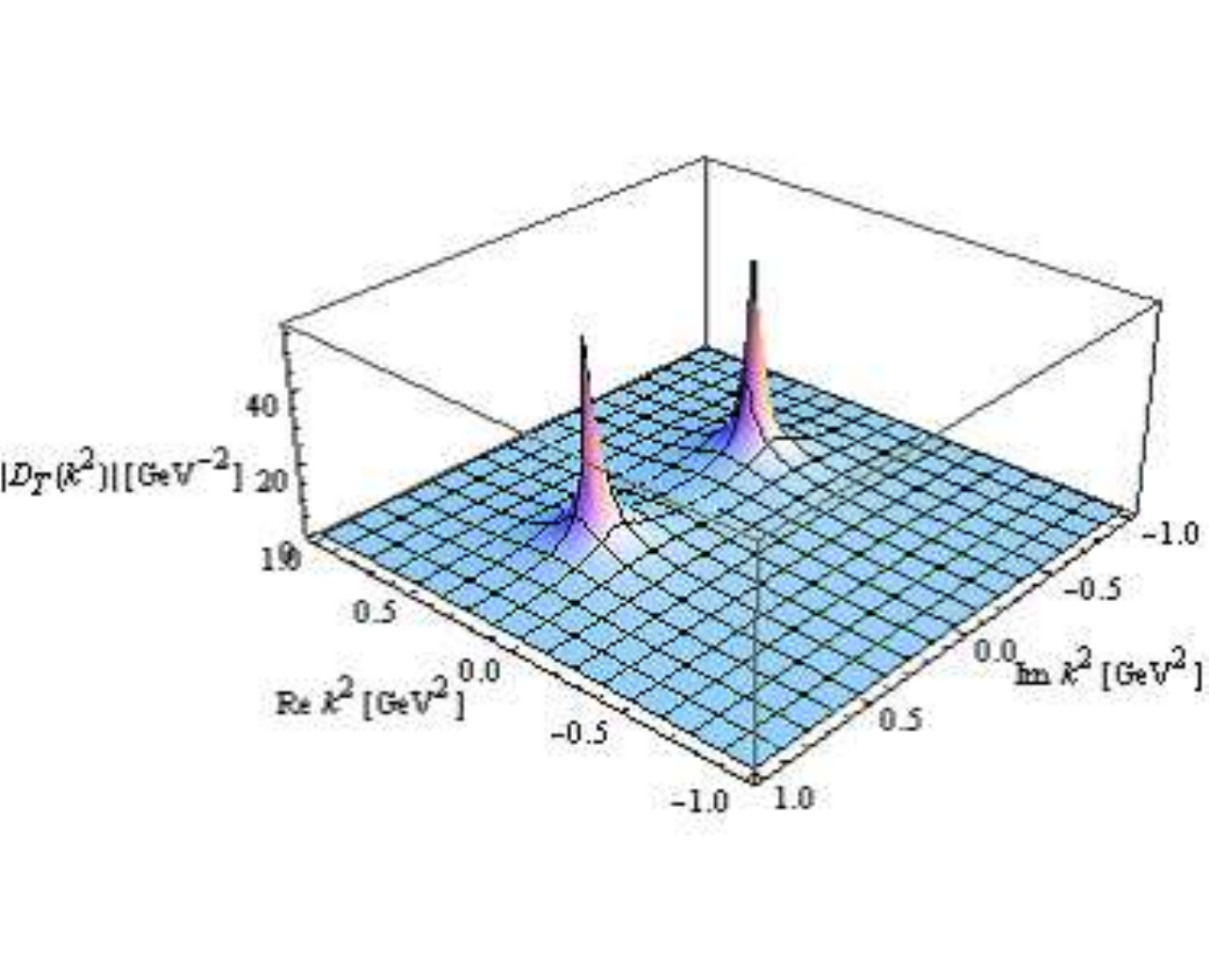}
 \end{center}
 \caption{Modulus of the gluon propagator $|{\mathscr D}_T(k^2)|$ at the value of (\ref{eq:physical_value}) with [TW] renormalization condition and $G = SU(3)$. The two spikes represent complex conjugate poles at $k^2 = 0.23 \pm 0.42i\ {\rm GeV^2}$}
 \label{3dpropagator.pdf}
\end{figure}

Here, we prove that the gluon propagator in the massive Yang-Mills model has two unphysical poles to one-loop order. The gluon propagator satisfies all the properties listed in the case II of Sec.~III.
 \begin{enumerate}
 \item ${\mathscr D}_T (z) \sim -\frac{1}{z} \tilde{D}(|z|)$ as $|z| \rightarrow \infty$, where $\tilde{D}(|z|)$ is a real function and positive for large $|z|$. Indeed, (\ref{TWgluonvacuumpolarization}) yields
 \begin{align}
{\mathscr D}_T(k^2) \simeq - \left[\left( \frac{g^2 C_2 (G)}{192\pi^2} \right) 26 k^2 \ln (|k|^2) + O(k^2)\right]^{-1}. \label{gluon_asymptotic_UV}
\end{align}
 \item $\rho(\sigma^2) < 0$ from (\ref{gluonnegativeity}).
 \item ${\mathscr D}_T(k^2 = 0) > 0$ from the assumption (\ref{pos_IR_prop_cond}).
 \end{enumerate}
Moreover, we find that the gluon propagator has no zeros, $N_Z = 0$, since $\Gamma_{{\mathscr A}}^{(2)}$ is finite from the expression (\ref{TWgluonvacuumpolarization}). According to (\ref{eq:case_2_number_of_complex_poles}), we conclude that the number of unphysical poles of the gluon propagator is two,
 \begin{align}
N_P = 2.
\end{align}
Therefore, the gluon propagator has a pair of complex conjugate poles or tachyonic poles with multiplicity two. For example, the modulus of the propagator $|{\mathscr D}_T(k^2)|$ on the complex $k^2$ at the value of (\ref{eq:physical_value}) with [TW] renormalization condition is plotted in Fig.~\ref{3dpropagator.pdf}, which shows a pair of complex conjugate poles at $k^2 = 0.23 \pm 0.42i\ {\rm GeV^2}$.

Note that we must take into account RG effects to reproduce the UV behavior of the Yang-Mills theory correctly, which is discussed in Sec.~VI B. We can verify that $N_P = 2$ holds for the infrared-safe parameter region within the one-loop RG improvement.

 Based upon the above results, we investigate the pole structure of the gluon propagator in the renormalization condition [TW]. Since we have proved $N_P = 2$ for the strict one-loop gluon propagator, it is sufficient to find a pole with a positive imaginary part as $k^2 = v+ iw, \ w \geq 0 $.
 
\begin{figure}[tbp]
 \begin{center}
 \includegraphics[width=70mm]{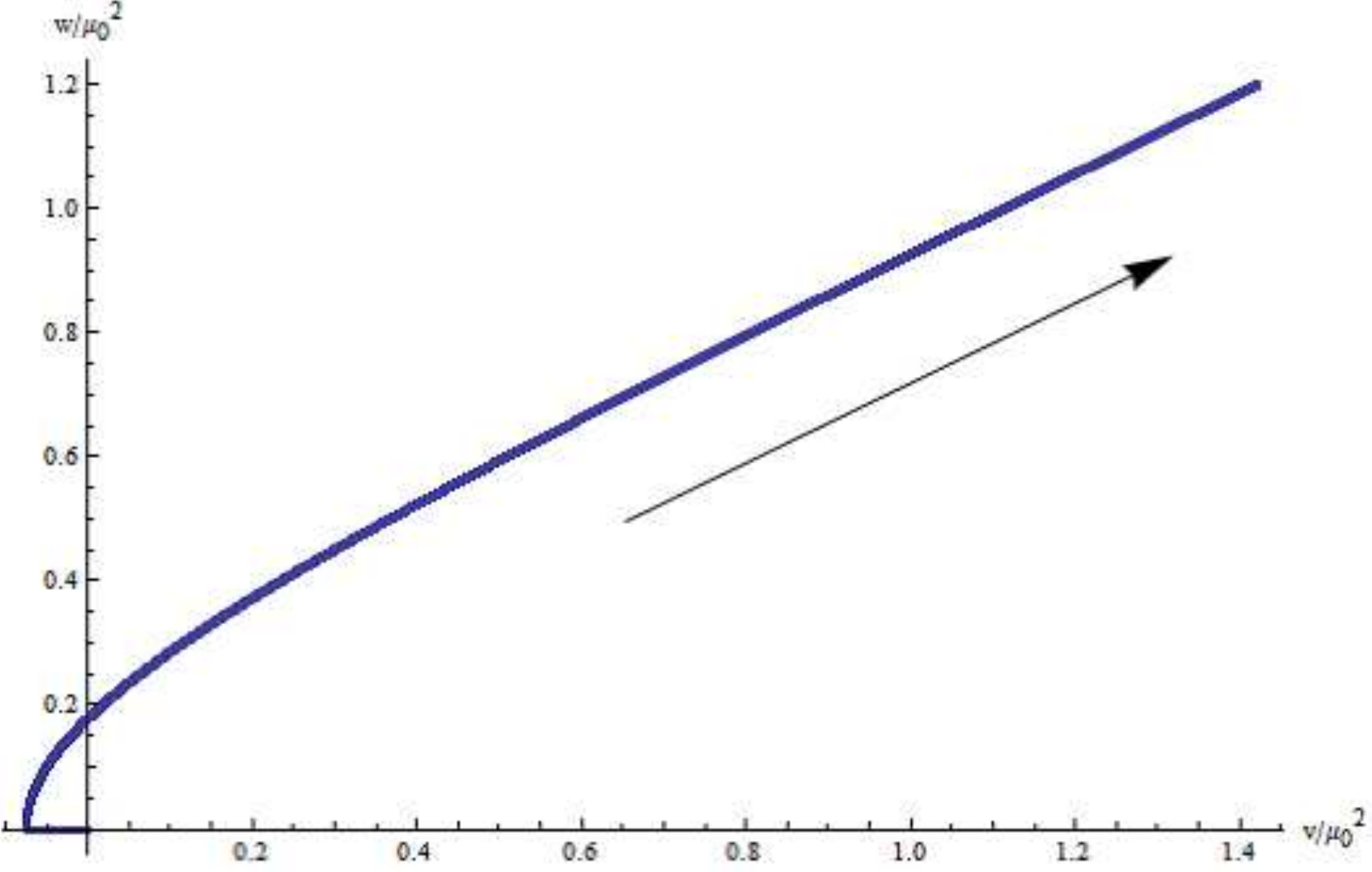}
 \end{center}
 \caption{Positions of poles at fixed $g = 4$ for varying $0 < M^2/\mu_0^2 < 1$ with [TW] renormalization condition and $G = SU(3)$. As $M^2$ increases, the position of the pole moves clockwise as shown by the arrow in the figure.}
 \label{poletraj.pdf}
\end{figure}

\begin{figure}[tbp]
 \begin{center}
 \includegraphics[width=70mm]{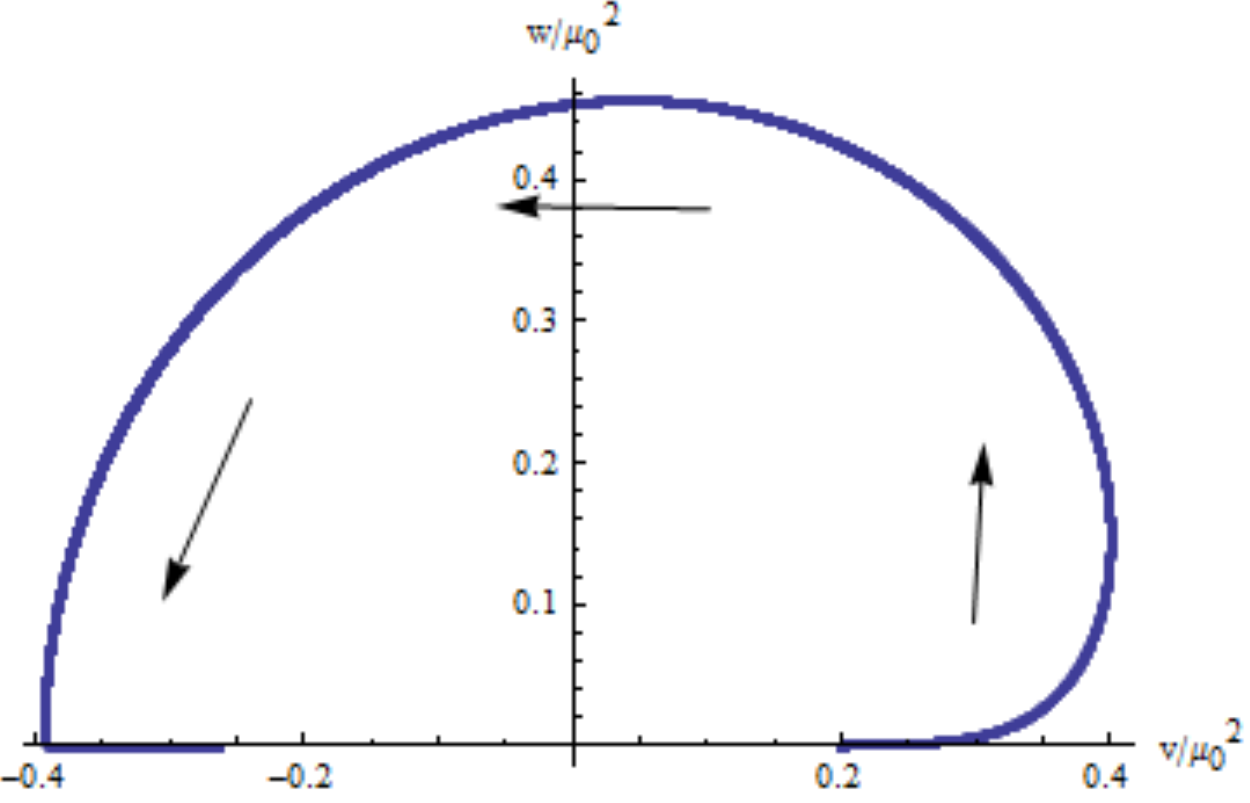}
 \end{center}
 \caption{Positions of poles at fixed $M^2/\mu_0^2 = 0.20$ for varying $0 < \lambda = \frac{C_2 (G) g^2}{16 \pi^2} < 3$ with [TW] renormalization condition and $G = SU(3)$. As $g^2$ increases, the position of the pole moves counterclockwise as shown by the arrows in the figure.}
 \label{poletraj_var_lambda_w_arrow.pdf}
\end{figure}

 Let us see how the pole position changes if one varies $M^2$ or $g^2$.
 Figure~\ref{poletraj.pdf} plots the pole positions of $k^2 = v + i w, \ w \geq 0$ at $g = 4$ for $0 < M^2/\mu_0^2 < 1$. As $M^2$ increases, tachyonic poles are transferred into a pair of complex conjugate poles, and the complex pole position $(v,w)$ grows, keeping the ratio $v/w \sim 1$.
 In addition, Fig.~\ref{poletraj_var_lambda_w_arrow.pdf} plots the pole positions of $k^2 = v + i w, \ w \geq 0$ at $M^2/\mu_0^2 = 0.20$ for $0<\lambda < 3$, where $\lambda := \frac{C_2(G) g^2}{16 \pi^2}$. As $g^2$ increases, the pole rotates counterclockwise and approaches the real negative axis to become a tachyonic pole. We will investigate the pole structure in more detail in Sec.~VI A.

 \subsection{Ghost propagator}
We can implement the similar analysis on the ghost propagator. We show that the ghost propagator does not have complex pole, and has at most one tachyonic pole, besides the massless pole. 
\subsubsection{Positivity of the ghost spectral function}
\begin{figure}[tbp]
 \begin{center}
    \includegraphics[width=70mm]{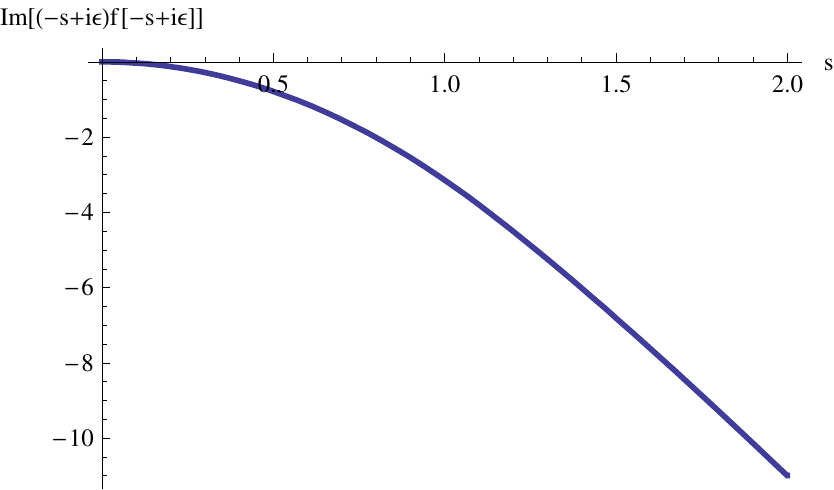}
 \end{center}
 \caption{Plot of ${\rm Im}\ [(- s + i\epsilon) f(-s + i \epsilon) ]$. This indicates $\rho_{gh}(\sigma^2) > 0$.}
 \label{ImPigh_for_article.pdf}
\end{figure}

We can show that the ghost spectral function $\rho_{gh}(\sigma^2) := \frac{1}{\pi} {\rm Im} \ \Delta(\sigma^2 + i \epsilon)$ is always positive for $\sigma^2 > 0$. Notice the negative sign in the definition of the ghost propagator (\ref{eq:ghost_def}). The dispersion relation yields
\begin{align}
\rho_{gh}(\sigma^2) > 0 \ \Leftrightarrow \ {\rm Im}\ \hat{\Pi}_{gh}(- \frac{\sigma^2}{M^2} + i \epsilon) < 0 .
\end{align}
Then, we find
\begin{align}
 {\rm Im}\ \hat{\Pi}_{gh}(- s + i \epsilon) = \frac{g^2 C_2 (G)}{64 \pi^2} {\rm Im}\ [ (-s+i\epsilon) f(-s+i\epsilon) ] < 0. \label{ghost_positivity}
\end{align}
See Fig.~\ref{ImPigh_for_article.pdf}. We deduce the positivity $\rho_{gh}(\sigma^2) >0$ for arbitrary $g^2>0, \ M^2$, irrespective of the renormalization condition.

\subsubsection{Absence of complex poles}
We shall apply the proposition of the case III in Sec.~III for $- \Delta(z)$. Note that the number of poles is not affected by an overall sign of the propagator.
 \begin{enumerate}
 \item $-\Delta (z) \sim - \frac{1}{z} \tilde{\Delta}(|z|)$ as $|z| \rightarrow \infty$, where $\tilde{\Delta}(|z|)$ is a real function and positive for large $|z|$. Indeed, the ghost propagator has the following asymptotic form from (\ref{TWghostvacuumpolarization})
 \begin{align}
-\Delta(k^2) \simeq - \left[\left( \frac{g^2 C_2 (G)}{64\pi^2} \right) 3 k^2 \ln (|k|^2) + O(k^2)\right]^{-1}.
\end{align}
 \item $-\rho_{gh}(\sigma^2) < 0$ for $\sigma^2 > 0$ from (\ref{ghost_positivity}).
 \item $-\Delta (z) \sim - Z_c/z$ with real $Z_c$ as $|z| \rightarrow 0$ as we can see that it has a pole at $k^2 = 0$.
 \end{enumerate}
Furthermore, the ghost propagator has no zeros, $N_Z = 0$, since $\Gamma_{gh}^{(2)}$ is finite from the expression (\ref{TWghostvacuumpolarization}).
According to (\ref{eq:case_3_complex_poles}), the ghost propagator has one tachyonic pole for $Z_c < 0$, and no unphysical pole for $Z_c > 0$. This is a result within the strict one-loop level, and we consider the RG improvement in Sec.~VI B, which leads to the same conclusion for $Z_c > 0$ in the infrared-safe parameter region.

If we take the [TW] renormalization condition, $Z_c$ becomes,
 \begin{align}
Z_c = \frac{1}{M^2 \left[ 1 + \frac{g^2 C_2 (G)}{64 \pi^2 } \left( \frac{5}{2} - f(\nu) \right) \right]}.
\end{align}
Note that $f(s)$ is a monotonically increasing function, $\left( \frac{5}{2} - f(\nu) \right) <0$. Therefore, the number of unphysical pole $N_{P,gh}$ reads
 \begin{align}
N_{P,gh} = \begin{cases}
0, \ \ g^2 < \frac{64 \pi^2}{C_2 (G) } \left( f(\nu) - \frac{5}{2} \right)^{-1} \\
1, \ \ g^2 > \frac{64 \pi^2}{C_2 (G) } \left( f(\nu) - \frac{5}{2} \right)^{-1}.
\end{cases}
\end{align}
The one unphysical pole must be a tachyonic pole, since complex poles must appear as a pair of complex conjugate poles. Hence the ghost propagator has no complex conjugate pole, and has (A) a massless pole with a negative norm or (B) a massless pole with a positive norm together with a tachyonic pole.

\subsection{Origin of complex poles}
The propositions in Sec.~III state that the negativity or quasinegativity of a spectral function yields the existence of complex poles. Let us mention origins of the negativity of the spectral function in this model. We consider the UV region and the IR region separately.

In the UV region, the asymptotic form of the function $h(s)$ for $|s| \rightarrow \infty$:
 \begin{align}
h(s) \rightarrow 26 \ln s
\end{align}
yields the spectral negativity. This term dominates ${\rm Im}\ [(-s+i\epsilon) h(-s+i\epsilon)]$ in $s \gtrsim 5$ . Notice that the negativity of the UV asymptotic expression of ${\rm Im}\ [(-s+i\epsilon) h(-s+i\epsilon)]$ can be identified with that of the anomalous dimension, which can be seen from (\ref{eq:dimregvacpol}). The asymptotic freedom, the negativity of the anomalous dimension and the RG analysis give the UV asymptotic form of the gluon propagator of the form
\begin{align}
D(k^2) \sim - \frac{Z_{UV}}{k^2 (\ln -k^2)^\gamma},
\end{align}
with $\gamma > 0$, which leads to the UV spectral negativity. Therefore, the negativity of the anomalous dimension plays an important role for the spectral negativity both in the strict one-loop level and in the RG-improved one in the UV region.

\begin{figure}[tbp]
 \begin{center}
    \includegraphics[width=70mm]{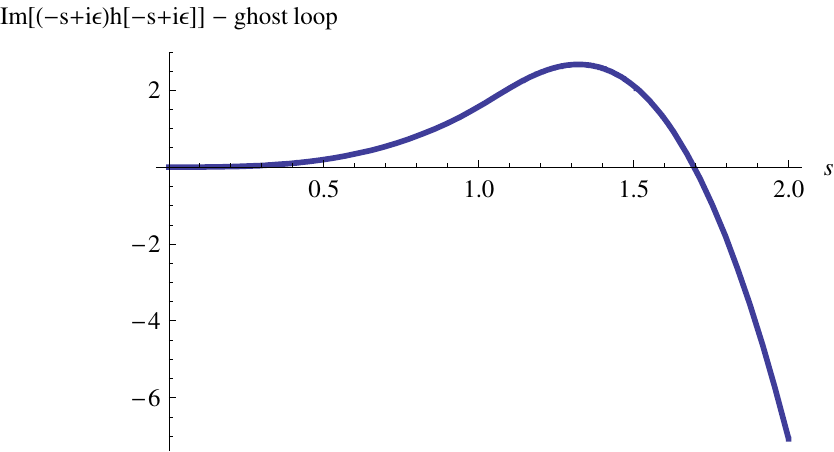}
 \end{center}
 \caption{Plot of ${\rm Im}\ [(- s + i\epsilon) h(-s + i \epsilon) ]$ with the ghost loop being subtracted. This indicates that the ghost loop contribution is essential for the negativity of the gluon spectral function $\hat{\rho}(\hat{\sigma}^2)$ in the IR region of $\hat{\sigma}^2 \lesssim 1.7$.}
 \label{fig:shs_without_ghosts.pdf}
\end{figure}

In the IR region, the $s \ln s$ term in (\ref{eq:dimregvacpol}), which arises from the ghost loop, makes the spectral function negative. Indeed, without the ghost loop, the dimensionless spectral function $\hat{\rho}(\hat{\sigma}^2)$ would be positive for $0<\hat{\sigma}^2<s_0$ and negative for $\hat{\sigma}^2>s_0$ where $s_0 \approx 1.7$ as shown in Fig.~\ref{fig:shs_without_ghosts.pdf}. In this sense, moreover, the masslessness of the ghost is important for the negativity of the gluon spectral function in the infrared. The ghost loop would not contribute to the spectral function up to $ 4 m_{gh}^2$, if the ghost had a mass $m_{gh}$.

The existence of complex poles in the one-loop massive Yang-Mills model arises from the spectral negativity in the light of the propositions in Sec.~III. The spectral negativity is due to the negativity of the anomalous dimension in the UV region, as in the pure Yang-Mills theory, and the ghost loop in the IR region.

\section{CONCLUSION and DISCUSSION}
Let us summarize our findings.
We have found that the modified superconvergence relation (\ref{eq:modified_superconvergence}) holds under the assumptions of one pair of complex conjugate poles (\ref{eq:one_pair_complex}) and the UV behavior (\ref{UVpropbehav}).
Notice that a violation of the usual superconvergence relation (\ref{eq:the_superconvergence}) implies existence of complex poles or other singularities if a propagator satisfies the UV behavior (\ref{UVpropbehav}).

We have derived the general relation between the winding number $(N_W(C))$ and the number of complex poles/zeros $(N_P/N_Z)$ of a propagator (\ref{eq:winding_number_relation}). As consequences of this equation, we deduce the value of $N_P- N_Z$, the difference between the number of unphysical poles and zeros, in the cases of quasipositive and quasinegative spectral functions.

We have applied this relation to the massive Yang-Mills model to find that the gluon propagator has two unphysical poles, and the ghost propagator has no complex poles. From the perspective of Sec.~III, the existence of complex poles of the gluon propagator stems from the negativity of the spectral function $\rho(\sigma^2)$, which arises from the ghost loop in the IR region, and the ordinary massless Yang-Mills theory framework in the UV region in the one-loop massive Yang-Mills model. Furthermore, we have examined the behavior of the pole position of the gluon propagator.

Let us add some comments on a very significant implication for gluon confinement that the results obtained in this paper would have.
It is well known that the gluon spectral function becomes negative in the ultraviolet region~\cite{OZ80a, OZ80b} in the Landau-gauge Yang-Mills theory.  Moreover, the negativity of the spectral function in a weak sense, namely the quasinegativity is enough to deduce the existence of complex poles. In the effective (massive Yang-Mills) model of the Landau-gauge Yang-Mills theory, indeed, we have confirmed that the gluon propagator has one pair of complex conjugate poles and no time-like pole.
These observations support that the gluon propagator in the Yang-Mills theory would have complex poles without a real pole.  The presence of complex poles invalidates the ordinary K\"all\'en-Lehmann spectral representation and therefore indicates the gluon confinement in the sense that the one gluon particle state must be excluded from the physical spectrum. Furthermore, the absence of a time-like pole in the gluon propagator also suggests that one gluon asymptotic state does not exist in the asymptotic state space in the Yang-Mills theory.

\subsection{Confinement/Higgs crossover}
We discuss the results of the pole structure in Sec.~V B, in relation to the confinement/Higgs crossover. In \cite{CFMPS16}, a phase transition to the Higgs branch as changing $M^2$ was reported in a similar model. On the other hand, Ref.~\cite{RSTW17} argues a smooth crossover instead of the phase transition as changing $M^2$ at $g = 4$ and $\mu_0 = 1$ GeV. We find no crossover to the Higgs region, based on the particlelike picture of the Higgs-like region.

To begin with, in all parameters except $g = 0$, the gluon propagator has no physical pole and two unphysical poles, which can be interpreted as that the theory is always in the confined phase. From the viewpoint of the similarity between the massive Yang-Mills model and the radially fixed gauge-scalar model \cite{Kondo18,KSOMH18}, this can be related to the Fradkin-Shenker continuity \cite{FS79,OS78,lattice-gauge-scalar-fund}.
In the single confined phase, we call a parameter region with particlelike gluons Higgs region, where the gluons are confined but exhibit a particlelike behavior.

To this end, we focus on the ratio $v/w$. 
 For instance, $v/w = +\infty$ corresponds to a physical pole, $v/w = 0$ to a pure imaginary pole. If a propagator has complex poles with $v/w \gg 1$, the propagator represents a particlelike resonance. Otherwise, the propagator provides no particlelike picture. This suggests that $v/w = 1$ can be a yardstick of the crossover between the Higgs-like region and the confining region.

\begin{figure}[tbp]
 \begin{center}
 \includegraphics[width=70mm]{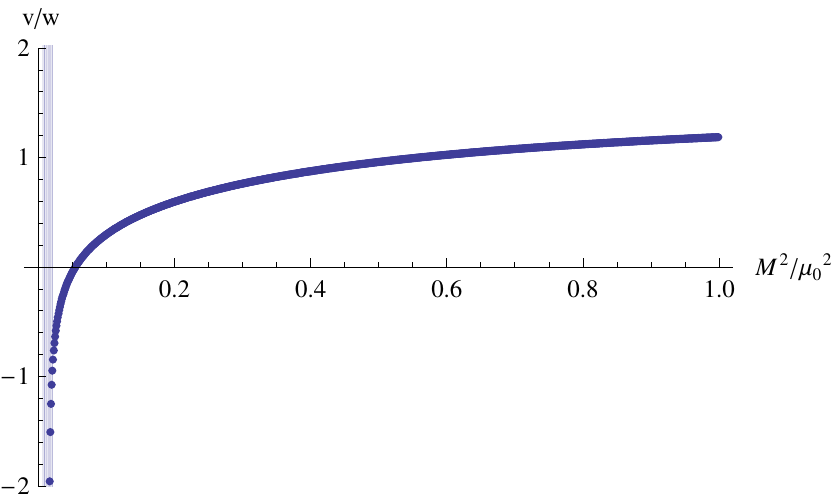}
 \includegraphics[width=70mm]{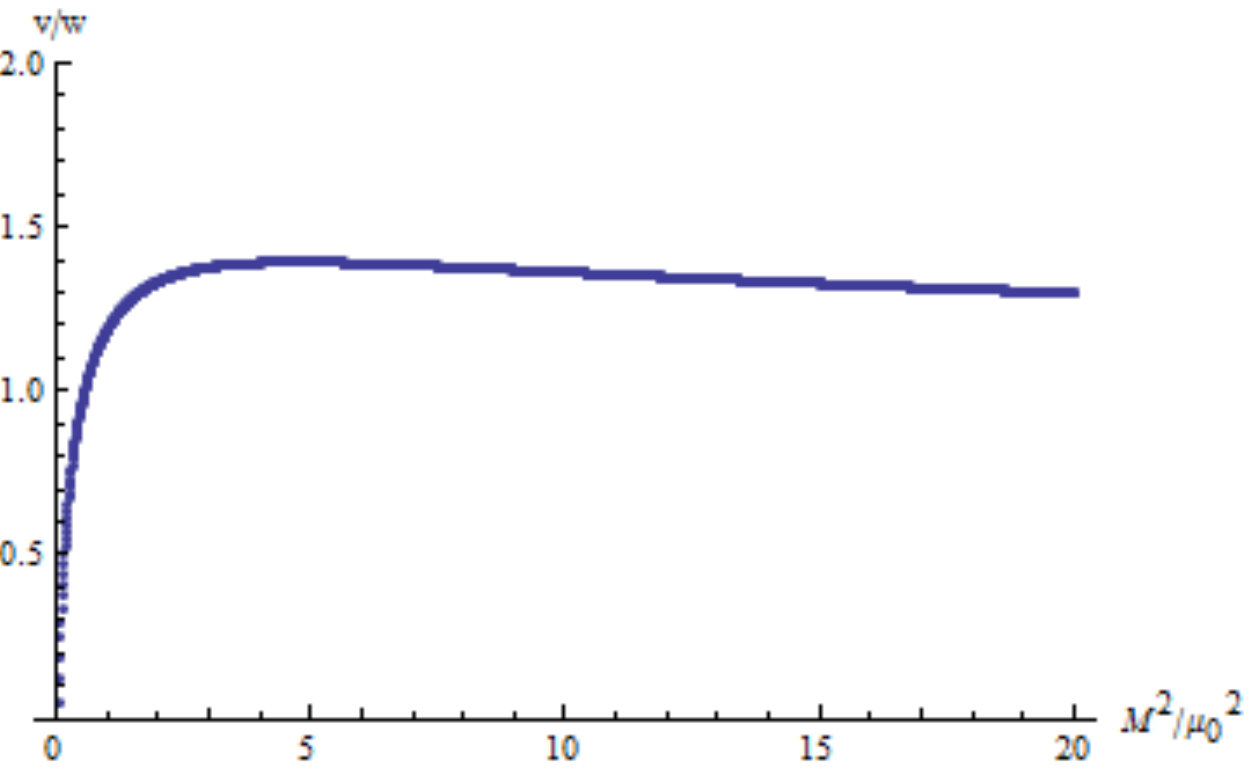}
 \end{center}
 \caption{Ratio $v/w$ for a pole of $k^2 = v + i w \ \ (w \geq 0)$ at fixed $g = 4$ with [TW] renormalization condition and $G = SU(3)$. A pair of pure imaginary poles appears at $M^2/\mu_0^2 = 0.54$. The existence of the finite upperbound on $v/w$ disproves the existence of the crossover to the Higgs region with the particlelike picture $v/w \gg 1$.}
 \label{w_v_Msq.pdf}
\end{figure}

\begin{figure}[tbp]
 \begin{center}
 \includegraphics[width=70mm]{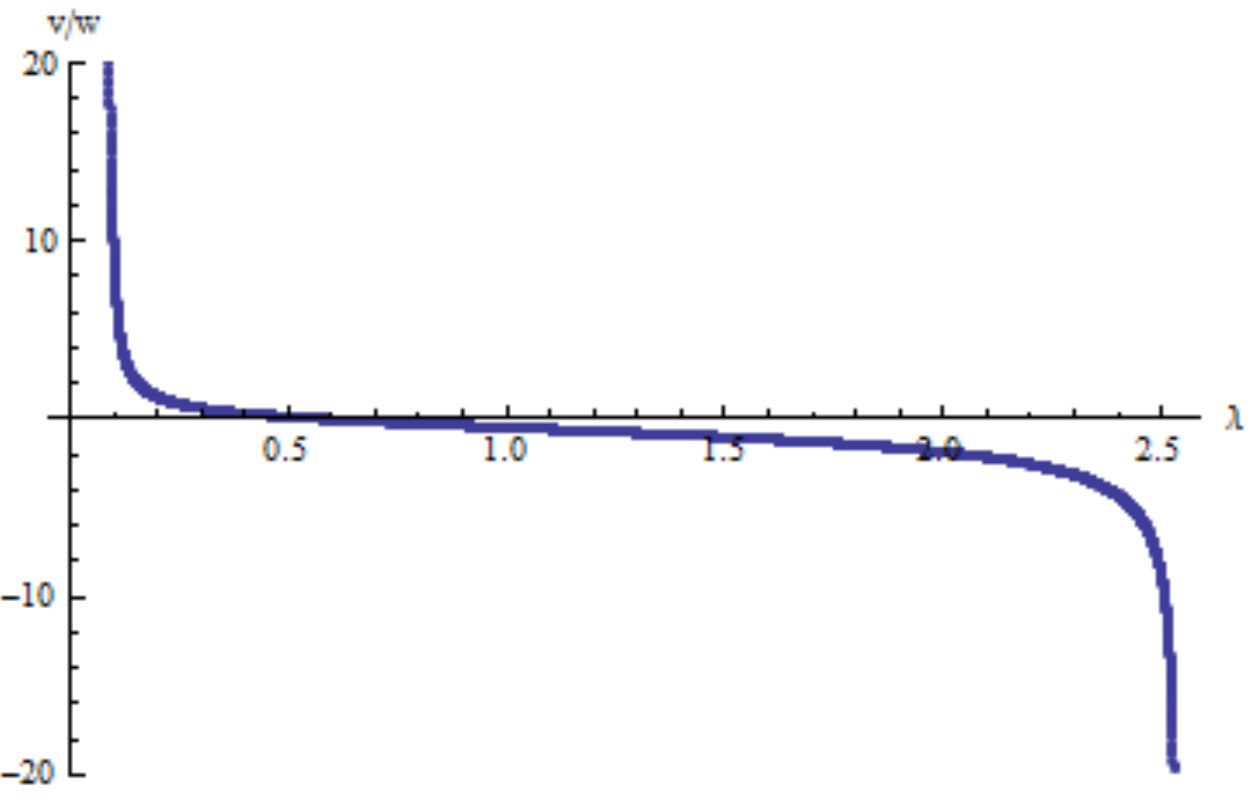}
 \end{center}
 \caption{Ratio $v/w$ for a pole for $k^2 = v + i w \ \ (w \geq 0)$ at fixed $M^2/\mu_0^2 = 0.20$ with [TW] renormalization condition and $G = SU(3)$, where $\lambda = \frac{C_2(G) g^2 }{16 \pi^2}$. A pair of complex conjugate poles is transferred into two tachyonic poles at $\lambda \approx 2.5$. This shows the confinement/Higgs crossover around $\lambda \sim 0.1$}
 \label{w_v_lambda.pdf}
\end{figure}

Figure~\ref{w_v_Msq.pdf} displays the ratio at $g = 4$ with various $M^2$.
This illustrates that $v/w \sim 1$ for large $M^2$ at $g = 4$, from which we find no clear evidence of the crossover to the Higgs-like region with the particlelike picture. Since large corrections from higher loops can arise around $M^2 \approx M^2_{scal}$, at which the RG-improved propagator is a scaling solution and the running coupling becomes large, a crossover between qualitatively distinct regimes can appear near $M^2 \approx M^2_{scal}$ in consistent with \cite{RSTW17}. We conclude that there may be qualitatively distinct regimes, but they are not the Higgs-like region with the particlelike gluons. On the other hand, Fig.~\ref{w_v_lambda.pdf} shows the ratio at $M^2/\mu_0^2 = 0.20$ with various $g^2$. This indicates the Higgs/confinement crossover in contrast to the case of changing $M^2$.
Notice that Fig.~\ref{w_v_Msq.pdf} and Fig.~\ref{w_v_lambda.pdf} correspond to Fig.~\ref{poletraj.pdf} and Fig.~\ref{poletraj_var_lambda_w_arrow.pdf} respectively.

%Ref.~\cite{RSTW17} reports the crossover appears around $M^2 \approx 2 M^2_{scal} \approx 0.146 \mu_0^2$ or $M^2 \approx 5 M^2_{scal} \approx 0.365 \mu_0^2$, where $M^2_{scal} \approx 0.073 \mu_0^2$ corresponds to the scaling solution in the RG improvement at $g=4$. Notice that a pair of pure imaginary poles appears at $M^2 = 0.054\mu_0^2$ near $M^2_{scal}$.

\begin{figure}[tbp]
 \begin{center}
    \includegraphics[width=70mm]{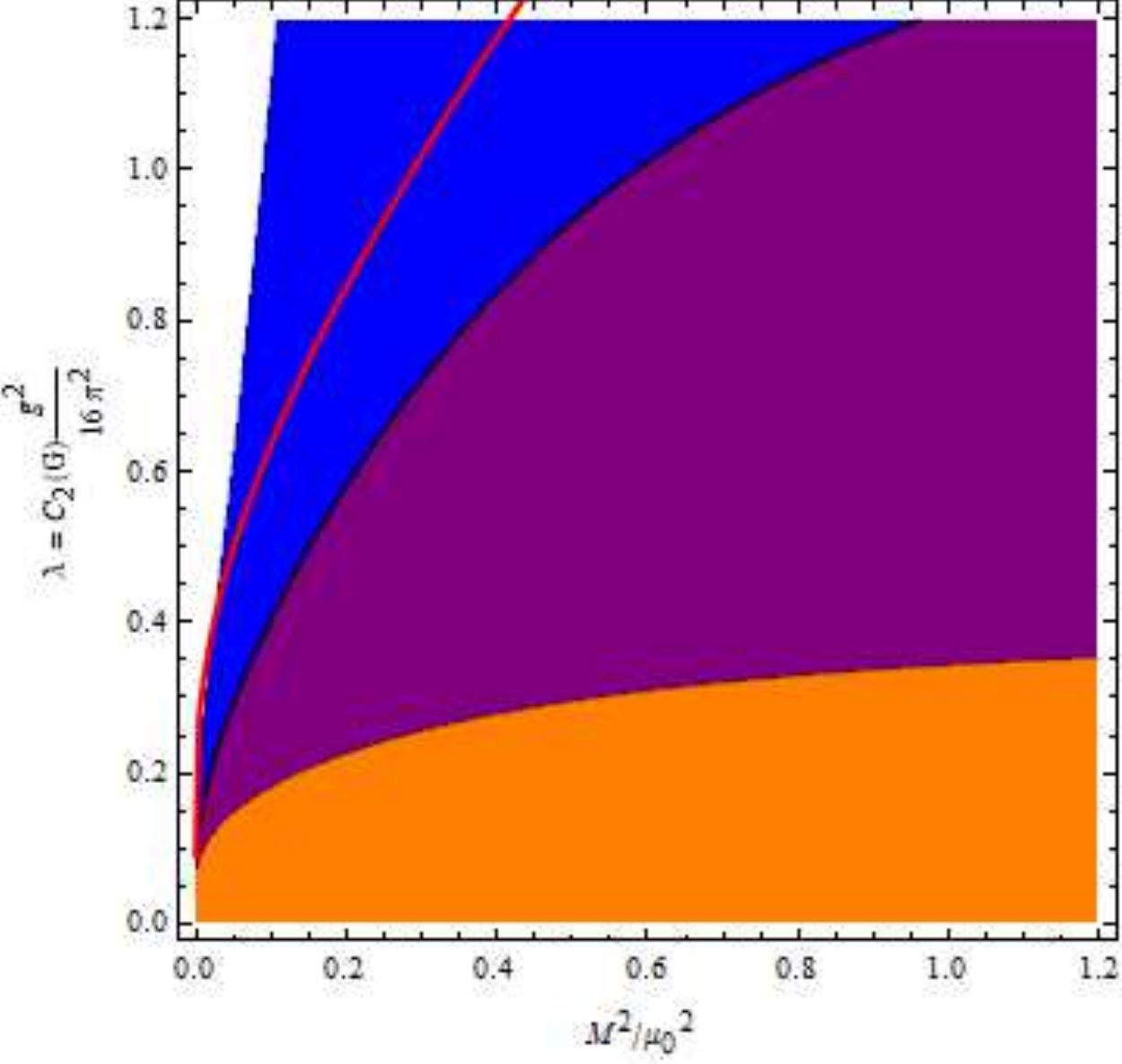}
    \includegraphics[width=70mm]{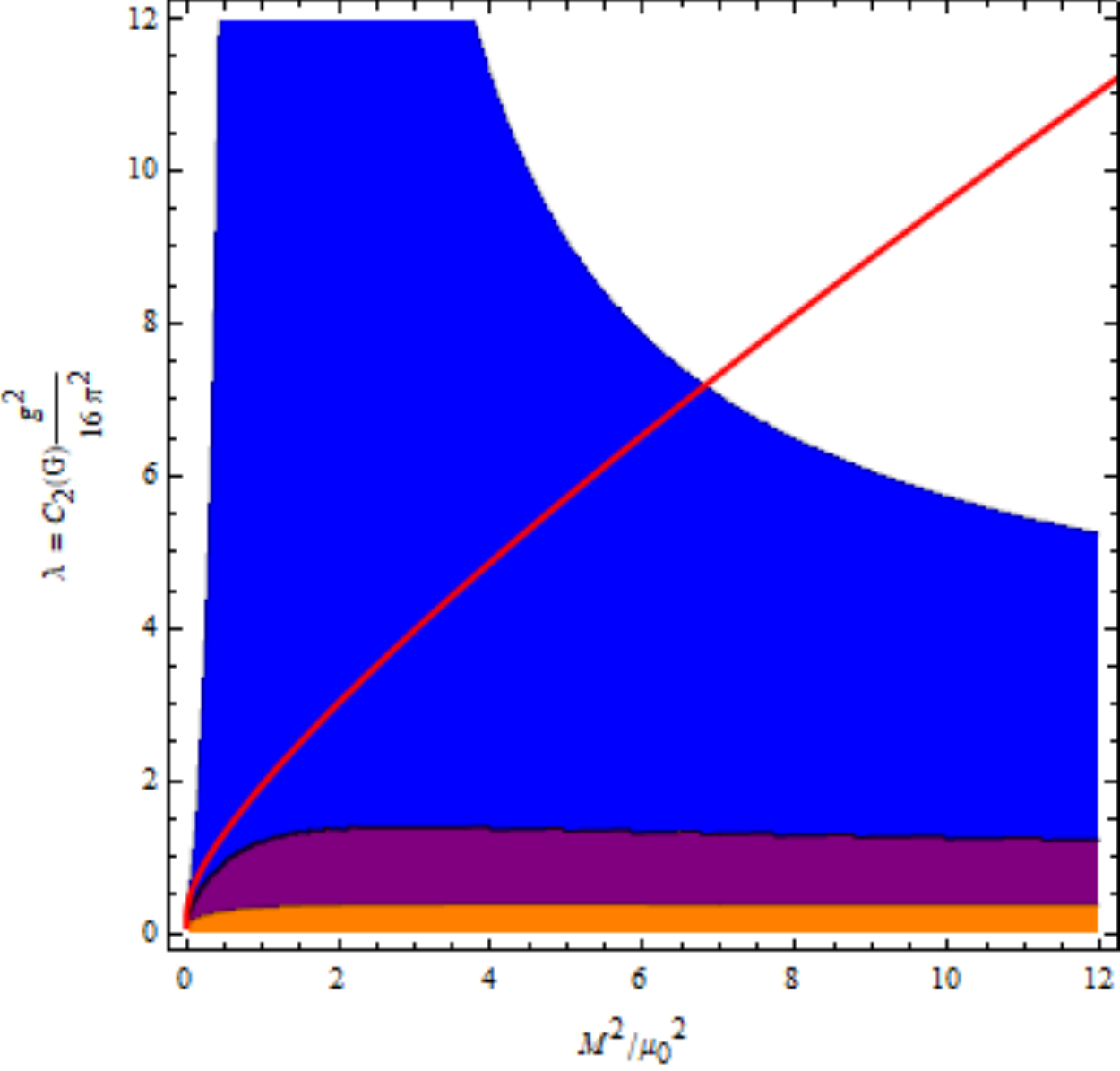}
 \end{center}
 \caption{Contour plots of $v/w$ for various parameters $(M^2,g^2)$ with [TW] renormalization condition and $G = SU(3)$. We color the region $v/w>1$ with orange, $1>v/w>0$ with purple, $0>v/w$ with blue and $w=0$ with white, signaling tachyonic poles. Above the red line of $g^2 = \frac{64 \pi^2}{C_2 (G) } \left( f(\nu) - \frac{5}{2} \right)^{-1}$, the ghost propagator has a tachyonic pole.}
 \label{fig:conf_higgs_crossover_contour.pdf}
\end{figure}

Finally, Fig.~\ref{fig:conf_higgs_crossover_contour.pdf} provides a schematic structure of the confining and Higgs-like regions in this model. We can expect the radially fixed gauge-scalar model can have a similar structure, from the equivalence of Ref.~\cite{Kondo18}.

Whereas these computations are done in the strict one-loop level, we can qualitatively justify the results in the parameter region where the running coupling is small, and $M^2, \mu_0^2$ and pole position are of the same order, as discussed below.

\subsection{On higher loop corrections and RG improvement}
The strict one-loop results get worse in higher energies, comparing with the results of numerical lattice simulations, due to large logarithms of the type $\ln \left( \frac{k^2}{\mu^2_0} \right)$. Therefore, we need to take into account RG effects to improve this situation. We discuss RG effects on the conditions to establish the number of complex poles.
 
 Let us see the general aspects of the RG improvement beforehand. The RG equation reads
 \begin{align}
&\Gamma_{{\mathscr A}}^{(2)}(k^2, \mu^2, g^2(\mu^2), M^2(\mu^2)) \notag \\
&~~~ = z_A (\mu^2,\mu_0^2) \Gamma_{{\mathscr A}}^{(2)}(k^2, \mu_0^2, g^2(\mu_0^2), M^2(\mu_0^2)),
\end{align}
 where $g(\mu^2)$ and $M(\mu^2)$ are the running coupling and mass and $z_A (\mu^2,\mu_0^2)$ is the factor of the anomalous dimension. In order to avoid the large logarithms, we can utilize this equation to obtain
 \begin{align}
&\Gamma_{{\mathscr A}}^{(2)}(k^2, \mu_0^2, g^2(\mu_0^2), M^2(\mu_0^2)) \notag \\
&~~~\approx [z_A (|k^2|,\mu_0^2) ]^{-1} \Gamma_{{\mathscr A},{\rm 1-loop}}^{(2)}(k^2, |k^2|, g^2(|k^2|), M^2(|k^2|)),
\end{align}
where $z_A (|k^2|,\mu_0^2)$ is calculated by using the one-loop anomalous dimension~ \cite{TW11}.
As reported in~\cite{TW11}, the running coupling used to fit the numerical results is small in all scales. Therefore, this approximation is reliable to some extent. Notice that it is the approximation that spoils the analyticity. Supposing that the exact propagator is analytic except for some points and the real positive branch cut, we shall evaluate the conditions to deduce the number of complex poles in this approximation under the assumption that the running coupling has no Landau pole. 
 \begin{enumerate}
 \item The asymptotic freedom yields
 \begin{align}
\Gamma_{{\mathscr A}}^{(2)}&(k^2, \mu_0^2, g^2(\mu_0^2), M^2(\mu_0^2)) \notag \\
&\sim -k^2 [z_A (|k^2|,\mu_0^2) ]^{-1}.
\end{align}
Therefore, although the RG improvement affects the power of the logarithm, the first condition ${\mathscr D}_T (z) \sim -\frac{1}{z} \tilde{D}(|z|)$ is maintained.
 \item We observe that ${\rm Im} \ \Gamma_{\mathscr{A},{\rm 1-loop}}^{(2)} (k^2 + i \epsilon,\mu^2, g^2, M^2 ) > 0$ for any $\mu$, $M$, and $g > 0$. Therefore, the second condition $\rho(\sigma^2) < 0$ still holds, since $z_A (|k^2|,\mu_0^2)$ does not change the sign.
 \item The RG-improved propagator satisfies the condition ${\mathscr D}_T (0) >0$ in the infrared-safe parameter region \cite{RSTW17}. Furthermore, although the propagator becomes the scaling solution on the separatrix between the infrared-safe and the Landau pole region, we can also establish $N_P =2$ for a scaling-type gluon propagator with a negative spectral function as discussed in Appendix A.
 \end{enumerate}
 Finally, ${\mathscr D}_T (k^2)$ has no zeros for $k^2 \neq 0$, as $\Gamma^{(2)}_{{\mathscr A},{\rm 1-loop}} (k^2, \mu^2, g^2, M^2 )$ is finite for any $k^2$, $\mu$, $g$, and $M$.
 Hence, the improved propagator satisfies the conditions in the infrared-safe region, which leads to $N_P = 2$. Note that this improvement excludes the case of tachyonic poles, since the improved two-point function in the Euclidean momentum $(k^2 < 0)$ reads $\Gamma^{(2)}_{{\mathscr A}} (k^2 ) = [z_A (|k^2|,\mu_0^2) ]^{-1} ( - k^2 + M^2 (|k^2|))$ because of the renormalization condition [TW], and is not zero.

 In addition, noting that the ghost spectral function $\rho_{gh} (\sigma^2)$ is positive for $\sigma^2 > 0$ in the one-loop level, and the RG improved ghost propagator behaves $-\Delta_{gh}(k^2) \sim - Z_c/k^2$ with $Z_c > 0$ as $|k^2| \rightarrow 0$ in the infrared-safe region, we similarly establish that the ghost propagator has no unphysical pole, $N_P = 0$, besides the massless pole with a negative-norm. Notice that the RG improvement also excludes the case of $Z_c < 0$. The appearance of tachyonic poles can be regarded as an artifact of the strict one-loop calculation.
 
Incidentally, it appears that no drastic correction on the above results of the pole positions is caused by higher loop corrections, if the pole position and the mass parameters $(M^2,\mu_0^2)$ are of the same order and the running coupling $g^2$ is small. If the running coupling is small in all scales, the one-loop RG improved propagator can be reliable, and therefore the strict one-loop propagator can be valid around $k^2 \approx \mu_0^2$. The flow diagram of the running coupling and mass is presented in Fig.~1 of Ref.~\cite{RSTW17}. In such a parameter region, the pole structure presented in Fig.~\ref{fig:conf_higgs_crossover_contour.pdf} is qualitatively valid. 

\subsection{Related topics and future works}

\begin{figure}[tbp]
 \begin{center}
    \includegraphics[width=70mm]{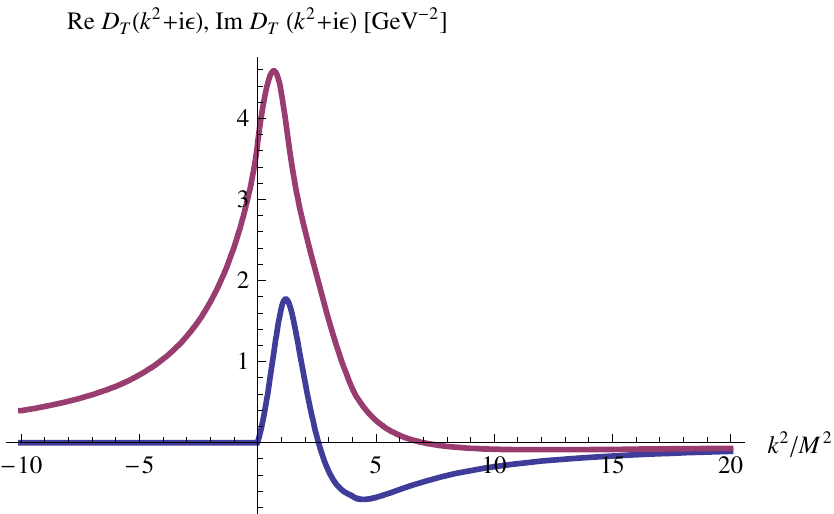}
 \end{center}
 \caption{Real part (red line) and imaginary part (blue line) of the gluon propagator for $N_f = 2$ at $g = 4,~ M^2/\mu_0^2 = 0.2$ with [TW] renormalization condition and $G=SU(3)$. This shows that the spectral function is quasinegative. Notice that the positivity of the spectral function in the IR region stems from the masslessness of the quark.}
 \label{fig:with2quarks}
\end{figure}

We have so far investigated the gluon propagator of the Yang-Mills theory, i.e., the quenched gluon propagator. Let us briefly comment on a quark-loop contribution to the gluon propagator. As reported in \cite{PTW14}, the quark loop contribution to the gluon vacuum polarization $\Pi_q(k^2)$ in [TW] renormalization condition reads
 \begin{align}
\Pi_q(k^2) &= \frac{g^2 C_2(r)}{6 \pi^2} k^2 \left[ h_q \left( \frac{m_q^2}{k^2} \right) - h_q \left( \frac{m_q^2}{\mu^2} \right) \right], \notag \\
h_q(t) &:= 2t + (1-2t) \sqrt{4t + 1} \coth^{-1} (\sqrt{4t+1}),
\end{align}
where $m_q$ is the quark mass and $C_2(r)$ is the quadratic Casimir operator of a representation $r$ of the quark. Note that the RG argument~\cite{OZ80a,OZ80b} shows that  the gluon spectral function remains negative in the UV region for $C_2(r) < \frac{13}{8} C_2(G)$, but it becomes positive for $\frac{13}{8} C_2(G)<C_2(r)<\frac{11}{4} C_2(G)$ according to negativity and positivity of the leading term of the anomalous dimension respectively, provided that $C_2(r)<\frac{11}{4} C_2(G)$ ensuring the UV asymptotic freedom. We now consider $N_f$ flavors of massless quarks in the fundamental representation with $C_2(r) = N_f/2$, to see the effect of the quark loop.
Although the dynamical quark mass should be included as an effective model due to the chiral symmetry breaking, we use the massless quark as a test, since the massless quark loop causes more potent positive contribution to the gluon spectral function than massive one.
One can find that the gluon has a quasinegative spectral function for $N_f = 1, \cdots,9 < 2 \times \frac{13}{8} C_2(G)$ around the physical parameters (\ref{eq:physical_value}) with $C_2(G)=3$ for $G=SU(3)$.
The real and imaginary parts of the gluon propagator for $N_f = 2$ are depicted in Fig.~\ref{fig:with2quarks}.
This suggests the inclusion of quarks with suitable parameters and $C_2(r) < \frac{13}{8} C_2(G)$ should not change the number of complex poles. Detailed analyses, including the quark propagator, remain for future work.

The massive Yang-Mills model can also be an effective theory at finite temperature.
An explicit expression of the gluon propagator of the massive Yang-Mills model at finite temperature can be found in Ref.~\cite{RSTW14-17}. 
Reference~\cite{Siringo17} presents the pole trajectory in the long-wavelength limit with varying $T$ in a similar model and reports a crossover of the pole location near the deconfinement temperature.
It would be interesting to consider the pole structure and the confinement/Higgs crossover in the finite-temperature, especially in relation to the deconfinement transition, and a comparison with phenomenological quasiparticle models of massive gluon~\cite{quasi-particle_massive_gluon}.

It would also be interesting to discuss a kinematic origin of complex poles in the massive Yang-Mills model, which violates the K\"all\'en-Lehmann spectral representation.

\section*{Acknowledgements}
This work was supported by Grant-in-Aid for Scientific Research, JSPS KAKENHI Grant Number
(C) No.15K05042.

\appendix

 \section{Various cases of the relation between unphysical poles and winding number}
 In addition to Sec.~III, we consider more examples of propagators including the scaling-type infrared behavior. The evaluations of the winding number are essentially the same as those of Sec.~III.
 
 \subsection{Gluon propagator}
 We consider a propagator $D(z)$ satisfying the following conditions: 
 \begin{enumerate}
 \item The propagator has the leading asymptotic behavior: $D(z) \sim -\frac{1}{z} \tilde{D}(z)$ as $|z| \rightarrow \infty$, where $\tilde{D}(z)$ is a real and positive function $\tilde{D}(z)>0$ for large $|z|$.
 \item The spectral function $\rho(\sigma^2)$ is quasipositive or quasinegative.
 \item $D(k^2) \rightarrow Z (-k^2)^{\alpha-1}$, where $
 Z > 0$ and $2 > \alpha > 1$.
 \end{enumerate}
Then we obtain,
 \begin{align}
N_Z - N_P = 
\begin{cases}
0 \ \ &(1<\alpha<1.5, \ \rho\mbox{ is quasipositive}) \\
-2 \ \ &(1<\alpha<1.5, \ \rho \mbox{ is quasinegative}) \\
-2 \ \ &(1.5<\alpha<2, \ \rho\mbox{ is quasipositive}) \\
-2 \ \ &(1.5<\alpha<2, \ \rho \mbox{ is quasinegative}). \label{eq:appendix_gluon_winding_number}
\end{cases}
\end{align}
We can evaluate $N_Z-N_P$ as in Sec.~III.
Note that as $k^2 \rightarrow +0$, $\frac{D(k^2 - i \epsilon)}{|D(k^2 - i \epsilon)|} = e^{i(\alpha - 1) \pi}$. For $1<\alpha<1.5$, the evaluation is basically the same as the case I' in Sec.~III.
On the other hand, the winding number of a quasipositive spectral function for $1.5<\alpha<2$ is different from that of $1<\alpha<1.5$. For simplicity, we now focus on a phase trajectory of the lower line on the real axis $(+ \epsilon, \infty)$ shown in Fig.~\ref{fig:ghost_complex_plane.pdf}, which has the initial value $D/|D| = -1$ and the final value $D/|D| = e^{i(\alpha - 1) \pi}$. For $1.5<\alpha<2$, a phase trajectory with a quasipositive spectral function, which does not pass through at $D/|D| = + i$ in the lower line, rotates clockwise in total, while it rotates counterclockwise in total for $1<\alpha<1.5$. Therefore, for $1.5<\alpha<2$, the winding number of a propagator with a quasipositive spectral function is the same as that of a quasinegative case. Then, we can calculate the winding number in each case similarly, and obtain
(\ref{eq:appendix_gluon_winding_number}).

 \subsection{Ghost propagator}
 Next, let us examine the following propagator $D(z)$: 
 \begin{enumerate}
 \item The propagator has the leading asymptotic behavior: $D(z) \sim -\frac{1}{z} \tilde{D}(z)$ as $|z| \rightarrow \infty$, where $\tilde{D}(z)$ is a real and positive function $\tilde{D}(z)>0$ for large $|z|$.
 \item The spectral function $\rho(\sigma^2)$ is quasipositive or quasinegative.
 \item $D(k^2) \rightarrow Z (-k^2)^{\beta-1}$, where $
 Z > 0$ and $0 > \beta > -1$.
 \end{enumerate}
Similarly, we have
 \begin{align}
N_Z - N_P = 
\begin{cases}
0 \ \ &(-0.5<\beta<0, \ \rho\mbox{ is quasipositive}), \\
0 \ \ &(-0.5<\beta<0, \ \rho \mbox{ is quasinegative}), \\
+2 \ \ &(-1<\beta<-0.5, \ \rho\mbox{ is quasipositive}), \\
0 \ \ &(-1<\beta<-0.5, \ \rho \mbox{ is quasinegative}).
\end{cases}
\end{align}

\section{Analyses on other models}
The consequences of Sec.~III can be applied to many models. We see consistencies with the various results in the literature.
\subsection{Massive expansion model}
\begin{figure}[tbp]
 \begin{center}
  \includegraphics[width=70mm]{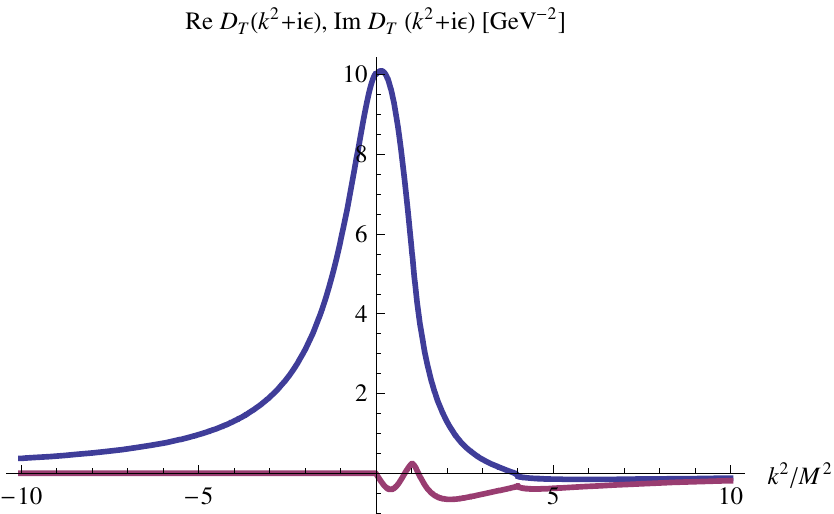}
 \end{center}
 \caption{The gluon propagator $\mathscr{D}_T (k^2+i\epsilon)$ in the model presented in \cite{Siringo16a, Siringo16b}. The blue line plots ${\rm Re}\ \mathscr{D}_T(k^2+i\epsilon)$ and the red line plots ${\rm Im}\ \mathscr{D}_T(k^2+i\epsilon) = \pi \rho (\sigma^2)$. This indicates that the spectral function $\rho$ is quasinegative.}
 \label{fig:Siringo.pdf}
\end{figure}

In Ref.~\cite{Siringo16a}, a double expansion for the Yang-Mills theory from the free massive Yang-Mills model in three-vertex, one-loop level, provides a following gluon propagator,
 \begin{align}
\mathscr{D}_T (k^2) &= \frac{1}{(- k^2)} \frac{1}{Z_F (F(-\frac{k^2}{M^2}) + F_0)}, \notag \\
F(s) &= \frac{5}{8 s} + \frac{1}{72} \Biggl[-s^2+\frac{2}{s^2}+2+\left(2-3 s^2\right) \ln (s) \notag \\
& ~~~ -\frac{(s+4)
 \left(s^2-20 s+12\right)}{s} +\frac{1}{s} \sqrt{\frac{s+4}{s}}\notag \\
& ~~~ \times \left(3
 s^3-34 s^2-28 s-24\right) \ln
 \left(\frac{\sqrt{s+4}-\sqrt{s}}{\sqrt{s}+\sqrt{s+4}}\right) \notag \\ 
 & ~~~ +\frac{2 (s+1)^2 \left(s^2-10 s+1\right)}{s^2} \notag \\
 &~~~+2 \frac{1}{s^3}
 (s+1)^2 \left(3 s^3-20 s^2+11 s-2\right) \ln
 (s+1) \Biggr]
\end{align}
where $G = SU(3)$, $M=0.73$ GeV, $Z_F = 0.30$, $F_0 = -1.05$. Figure~\ref{fig:Siringo.pdf} displays ${\rm Re}\ \mathscr{D}_T(k^2+i\epsilon)$ and ${\rm Im}\ \mathscr{D}_T(k^2+i\epsilon)$ and demonstrates that the spectral function is quasinegative. Since the two-point vertex function has no pole, we conclude that the number of unphysical poles is two. $N_P=2$ is consistent with \cite{Siringo17a}, which reports that the gluon propagator has a pair of complex conjugate poles.

 \begin{figure}[tbp]
 \begin{center}
  \includegraphics[width=70mm]{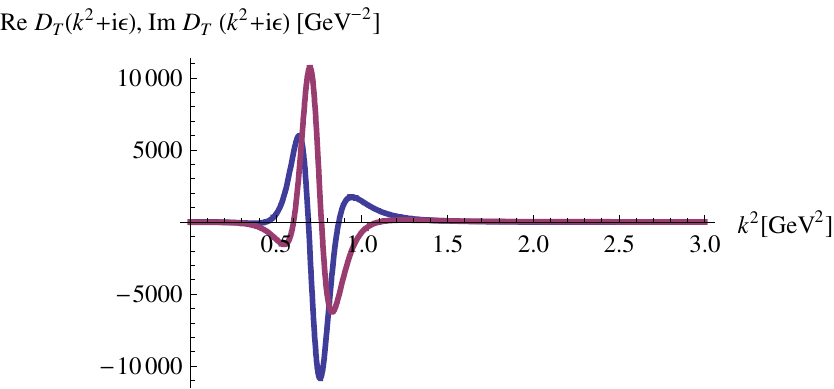}
  \includegraphics[width=70mm]{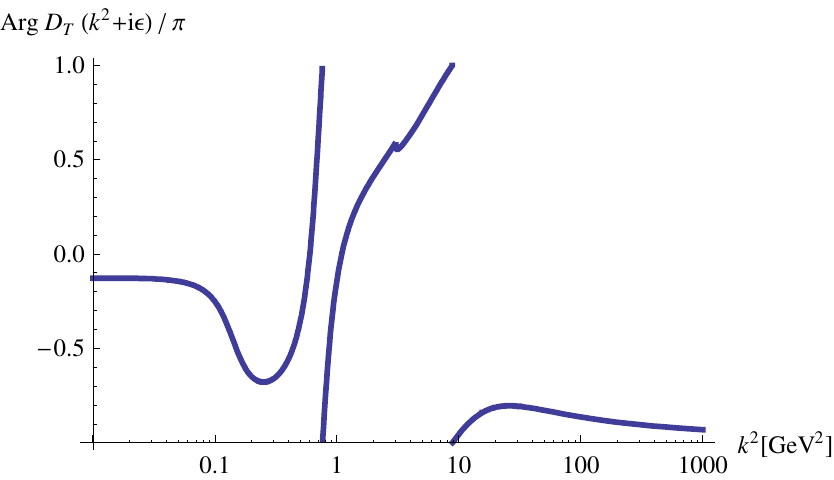}
 \end{center}
 \caption{The gluon propagator $\mathscr{D}_T (k^2+i\epsilon)$ obtained by the reconstruction \cite{CPRW18}. In the upper panel, the blue line plots ${\rm Re}\ \mathscr{D}_T(k^2+i\epsilon)$ and the red line plots ${\rm Im}\ \mathscr{D}_T(k^2+i\epsilon)$. The lower panel shows the principal value of the argument of $\mathscr{D}_T (k^2+i\epsilon)$. This plot and $\mathscr{D}_T (k^2) \rightarrow (-k^2)^{0.134641}$ as $k^2 \rightarrow 0$, which guarantees no winding number around the origin, indicate $N_W (C_2) \approx + 3$.}
 \label{fig:Cyrolgluon}
\end{figure}

\subsection{Numerical solving DSEs in the complex momentum plane}
In Ref.~\cite{SFK12}, the gluon propagator on the complex momentum plane of the Yang-Mills theory was computed by means of the truncated Dyson-Schwinger equations (DSEs). There, it was reported that the gluon propagator has no complex pole, ${\rm Re} \ \mathscr{D}_T (p_0^2) = 0$ where $p_0^2 \approx 0.3 $ GeV$^2$ and the gluon spectral function is positive there, $\rho(p_0^2) > 0$ from Fig.~5 in Ref.~\cite{SFK12}. Therefore, the quasipositivity of the gluon spectral function yields $N_P - N_Z = 0$, which can be consistent with the reported result $N_P = 0$.
\\
\subsection{A reconstruction of the gluon propagator}
In Ref.~\cite{CPRW18}, the QCD gluon propagator was constructed by using the ansatz consisting of the generalized Breit-Wigner basis, a polynomial, a power, and a logarithmic factor, based on the data from the FRG approach \cite{CFMPS16},
 \begin{align}
\mathscr{D}_T (k^2) &= {\cal K} \hat{G}(-k^2/M^2), \notag \\
\hat{G}(s) &= s^{-1-2\alpha} \left[ \ln \left( 1+ \frac{s}{\hat{\lambda}^2} \right) \right]^{-1-\beta} \notag \\
&\times \prod_{j=1}^6 \left[ \frac{\hat{{\cal N}}_1}{(\sqrt{s} + \hat{\Gamma}_{1,j})^2 + M_{1,j}^2} \right]^{\delta_{1,j}} \left[ \sum_{k=1}^5 \hat{a}_k s^{k/2} \right],
\end{align}
where ${\cal K} = 1$ GeV$^{-2}$, $M = 1$ GeV and other parameters are given in Table I of Ref.~\cite{CPRW18}. Note that the Breit-Wigner propagator does not have complex poles on the usual complex $k^2$ sheet and have poles only ``inside the branch cut'', which leads to $N_P = 0$. On the other hand, the polynomial term $\sum_{k=1}^5 \hat{a}_k s^{k/2} $ yields two nontrivial zeros of $s \neq 0$ in the usual complex $k^2$ sheet when we choose the parameters given in Ref.~\cite{CPRW18}, from which $N_Z = 2$. Therefore, the winding number becomes $N_W(C) = +2$ in the reconstructed gluon propagator. This is an example, whose spectral function is neither quasipositive nor quasinegative, having a nontrivial winding number along the path on the real axis. 

The winding number $N_W(C) = +2$ can be directly checked by following the phase of the gluon propagator. Indeed, Fig.~\ref{fig:Cyrolgluon} and the fact that $\mathscr{D}_T (k^2) \rightarrow (-k^2)^{0.134641}$ as $k^2 \rightarrow 0$ show that $N_W (C_2) \approx + 3$. On the other hand, the reconstructed gluon has the UV asymptotic behavior $\mathscr{D}_T (k^2) \sim \frac{1}{(-k^2)^{1.011279} (\ln |k^2|)^{0.222787}}$, which provides $N_W(C_1) \approx -1 $. We therefore have the winding number $N_W(C) = +2$, which is consistent with the number of poles and zeros.

\end{document}